\documentclass[useAMS,usenatbib]{mn2e}

\usepackage{natbib}
\usepackage{graphicx}
\usepackage{subfigure}
\usepackage{upgreek}
\usepackage{color}
\usepackage{enumerate}
\usepackage{verbatim}  
\usepackage{booktabs}
\usepackage{amsmath}
\usepackage[table]{xcolor}
\usepackage[section]{placeins}
\usepackage{float}
\usepackage{bm}
\begin{document}
\title[Clumpy tori in AGNs]{Clumpy tori illuminated by the anisotropic radiation of accretion discs
in active galactic nuclei}
\author[J. J. He et al.]{Jian-Jian He$^{1,2}\thanks{E-mail:
hejianjian@ihep.ac.cn}$, Yuan Liu$^{1}\thanks{E-mail:
liuyuan@ihep.ac.cn}$,
Shuang-Nan Zhang$^{1,3}\thanks{E-mail:
zhangsn@ihep.ac.cn}$\\
$^{1}$ Key Laboratory of Particle Astrophysics, Institute of High Energy Physics, Chinese Academy of Sciences, Beijing 100049, China\\
$^{2}$ University of Chinese Academy of Sciences, Beijing 100049, China\\
$^{3}$ Space Science Division, National Astronomical Observatories of
China, Chinese Academy of Sciences, Beijing 100012, China\\}

\maketitle
\begin{abstract}
 In this paper, we try to explain the observed correlation between the covering factor (CF) of hot dust and the properties of active galactic nuclei (AGNs), e.g., the bolometric luminosity ($L_{\rm{bol}}$) and black hole mass ($M_{\rm{BH}}$). Combining the possible dust  distribution in the torus, the angular dependence of the radiation of the accretion disc, and the relation between the critical angle of torus and the Eddington ratio,  there are eight possible models investigated in our work. We fit the observed CF with these models to determine the parameters of them. As a result, clumpy torus models can generally explain the observed correlations of tori, while the smooth models fail to produce the required CFs. However, there is still significant scatter even for the best-fitting model, which is the combination of a clumpy torus illuminated by the anisotropic radiation of accretion disc in an AGN. Although some of the observed scatter  is  due to the uncertainties in measuring $L_{\rm{bol}}$ and $M_{\rm{BH}}$, other factors are required in more realistic model. The models examined in this paper are not necessary to be the physical model of tori. However, the reasonable assumptions  selected during this process should be helpful in constructing physical models of tori.
\end{abstract}


\section{Introduction}

The presence of a geometrically thick toroidal obscuration is pivotal to the unification scheme of active galactic nuclei (AGNs) \citep{Antonucci1993,Urry1995}. The central engine is surrounded by a dusty toroidal structure (commonly referred to as the torus) so that the observed diversity of AGNs simply reflects different viewing angles of an axisymmetric geometry. The first direct evidence for torus existence has been provided by \cite{Jaffe2004} in NGC 1068. Since then the properties of tori have been investigated comprehensively, e.g., geometry, column density distribution, and the composition of dust. A physical model of torus is important for the understanding of the feedback of AGNs, the accretion of black holes, and the co-evolution of black holes and galaxies.

In order to reproduce the observed dust emission in the IR, various radiative transfer models have been developed, which can be divided into two classes: (1) smooth models, in which the dust is smoothly distributed in the torus; and (2) clumpy models, in which the dust is assumed to be clumped into discrete clouds. These two classes of models are not in competition, but just the consequence of different approaches to the complex physics of AGNs.

The IR spectral energy distribution (SED) of smooth model is relatively easy to calculate; however, smooth models cannot work well to match the observed narrower IR bump \citep{Dullemond2005}. Although the $10\ {\mu m}$ silicate feature is observed in absorption in type II AGNs, but it is shallower than predicted \citep{Dullemond2005}. \cite{Nenkova2008} have suggested a model that for the first time takes proper account of the clumpy structure of the tori. They presented the general formalism for radiative transfer in clumpy media and showed that a large range of dust temperatures coexist at the same distance from the central source in the clumpy torus. The model has naturally explained the lack of deep $10\ {\mu m}$ absorption features in AGNs  and can reproduce the weak emission feature  detected in type II AGNs. However, the model cannot reproduce the near-infrared emission from the hot dust in type I AGNs \citep{Mor2009}, which is a general behaviour in other clumpy models \citep{Polletta2008}.

The current models can explain some aspects of the observed properties of tori. Nevertheless, extreme simplifications are still required in both clumpy and smooth models, e.g., a power-law distribution of the dust in radial direction is often involved \citep{Barvainis1987,Nenkova2008}. Since AGN unification requires geometrical thickness, the angular dependence of the distribution is also described in a simple form, e.g., a Gaussian distribution \citep{Nenkova2008}. These approaches are largely based on picking the `right' distribution of the dust that best fits the observations. More recently the hydrodynamic process and radiative feedback have been included in the modelling of the structure of tori. However, due to the technical difficulties, it cannot include all critical physics, e.g., self-consistent clumping and the effect of magnetic field \citep{Wada2012,Schartmann2014}.

In the observational aspect, significant correlations have been found between the dust covering factor (CF) and the properties of AGN, such as bolometric luminosity, black hole mass, and Eddington ratio \citep{Cao2005,Maiolino2007,Treister2008}.  Recently, the $Wide\ field\ Infrared\ Survey\ Explorer$ ($WISE$) has provided very rich data in near- and mid-infrared bands, which allow us to study the IR emission and the dust CF of AGNs with quite a large sample. The recent work of \cite{Mor2011}, who combined optical spectra from the Sloan Digital Sky Survey and MIR photometry from the preliminary data release of $WISE$, suggested that there is a strong anticorrelation between the hot dust covering factor ($\rm{CF_{\rm{HD}}}$) and the bolometric luminosity ($L_{\rm{bol}}$), a weak correlation between $\rm{CF_{\rm{HD}}}$ and black hole mass ($M_{\rm{BH}}$), and no dependence between $\rm{CF_{\rm{HD}}}$ and $L_{\rm{bol}}/L_{\rm{Edd}}$. Other works of \cite{Calderone2012}, \cite{Ma2013} and \cite{Roseboom2013} also find similar statistical properties of AGNs.

However, the physical mechanism responsible for the observed statistical properties is still not clear. The receding torus scenario \citep{Lawrence1991} or the model of \cite{Nenkova2008} cannot self-consistently predict the values of CF of the tori with the given $L_{\rm{bol}}$ and $M_{\rm{BH}}$. Meanwhile, it is necessary to improve the models and enhance our understanding of the geometry and composition of the hot dust. Here, we try to construct models to explain the observed statistical properties mentioned above, by combining the current understanding of the dust distribution, the accretion disc radiation, and the critical angle of tori. Because the physical models of tori are still in their infancy, we do not intend to build a physical model from first principles, but instead just test if the assumptions usually adopted in literature can explain the observed properties of tori. The comparison with the observational trends can pick out the more reasonable ones, in order to make progress towards constructing physical models of tori.

The plan of this work is as follows. The models we construct are summarized in Section 2. In Section 3, the comparison between the models and observed $\rm{CF_{\rm{HD}}}$ is presented. In Section 4, we discuss the implication for the future model of tori and give our conclusions. Throughout this work, we use the sample of \cite{Mor2011}.

\section{Models}
We will determine $\rm{CF_{\rm{HD}}}$ as the function of the bolometric luminosity ($L_{\rm{bol}}$) and black hole mass ($M_{\rm{BH}}$). Thus, the correlation between $\rm{CF_{\rm{HD}}}$ and Eddington ratio ($L_{\rm{bol}}/L_{\rm{Edd}}$) is only an induced result.

 In our models, the dust is distributed in the form of clumps or smoothly distributed in torus. We choose three different methods to obtain the critical angle of the tori. The first is the observed ratio of type I AGNs to type II AGNs \citep{Lu2010}, which is the common method to obtain critical angle of torus. The second is the theoretical equation determined by \cite{Liu2011}, which relates critical angle of torus with $L_{\rm{bol}}/L_{\rm{Edd}}$. The last is the result from numerical simulations \citep{Dor2012}. The observed flux is $F=\frac{Lf(\theta)}{4\uppi d^2}$, where $L$ is the total luminosity of the source, $d$ is the distance between the observer and the source. The accretion disc radiation is isotropic  $f(\theta)=1$ or anisotropic $f(\theta)={\rm{2}}\cos\theta$.

\subsection{Dust distribution}

\begin{figure}
\centering
\includegraphics[width=8cm]{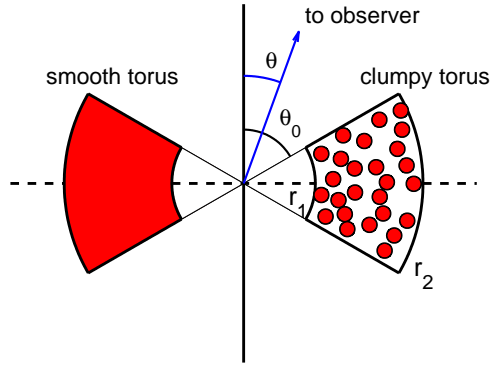}
\caption{The distribution of the dust in clumpy torus and smooth torus. The viewing angle of an observer is $\theta$, and  the critical angle of torus is $\theta_{0}$. $r_{1}$ is the evaporation radius and $r_{2}$ is the outer radius of torus.}
\label{Fig1}
\end{figure}

The distribution of the dust in torus is not clear yet. We have considered two forms of tori, which are shown in Fig.~\ref{Fig1}.

(1) Clumpy torus: in this case, the dust is assumed to be clumped into discrete clouds, which are approximated as uniform-density spheres of radius $l<<r_{\rm{d}}$, where $r_{\rm{d}}$ is the distance between two clumps. The radial number density of dusty clumps is
\begin{equation}
n_{\rm{cl}}=n_{\rm{cl,0}}\left(\frac{r}{r_{\rm{0}}}\right)^{-\alpha},
\end{equation}
where $n_{\rm{cl,0}}$ is the cloud number density at $r_0=0.5\ {\rm{pc}}$. All clouds have the same radius $l$.

(2) Smooth torus: here it is assumed that the dust is smoothly distributed and the number density of the dust is
\begin{equation}
n_{\rm{gr}}=n_{\rm{gr,0}}\left(\frac{r}{r_{\rm{0}}}\right)^{-\alpha},
\end{equation}
where $n_{\rm{gr,0}}$ is the grain number density at $r_0=0.5\ {\rm{pc}}$.

The form of the dust distribution is somewhat arbitrary; however, as shown in the appendix, our main conclusion is not sensitive to more complex form. A more realistic form will be considered in future works if it is well determined as a prior, e.g., by infrared interferometer.

\subsection{The relation between the critical angle ($\theta_{\rm{0}}$) of torus and Eddington ratio ($L_{\rm{bol}}/L_{\rm{Edd}}$)}

Both the gravitational force and radiative pressure work on the dust, when the gravitational force is balanced by the radiative pressure, we can obtain a critical angle $\theta_{\rm{0}}$ of torus, shown in Fig.~\ref{Fig1}. In Section 2.1, we have defined the number density of dust ($n_{\rm{cl}}$ and $n_{\rm{gr}}$). In order to obtain the CF of tori, we need to determine $\theta_0$ with the given $L_{\rm{bol}}$ and $M_{\rm{BH}}$. We choose three different methods to determine the critical angle of tori.

\begin{figure}
\centering
\includegraphics[width=8cm]{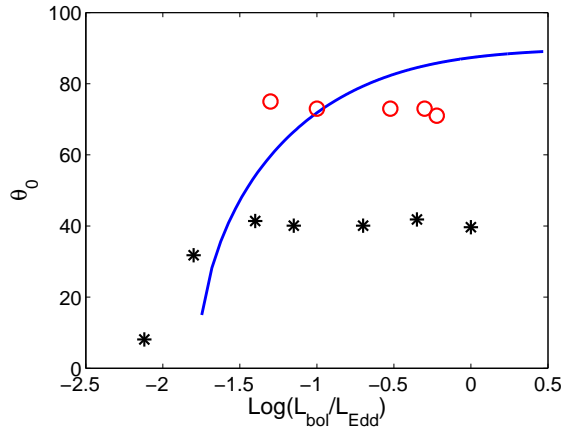}
\caption{The relations of the $\theta_{\rm{0}}\sim L_{\rm{bol}}/L_{\rm{Edd}}$: blue line, black stars and red circles are the relations given by Liu \& Zhang (2011), Lu et al. (2010) and Dorodnitsyn \&  Kallman (2012), respectively.}
\label{Fig2}
\end{figure}

(1) \cite{Liu2011} suggested a relation between  $\theta_{\rm{0}}$ and $L_{\rm{bol}}/L_{\rm{Edd}}$, i.e.,
\begin{equation}
\cos \theta_{\rm{0}}=\frac{\sqrt {\rm{1}+\rm{28}/(\rm{3}\mathnormal{\lambda})}-1}{\rm{4}},
\end{equation}
where $\mathnormal{\lambda}=\mathnormal{A}L_{\rm{bol}}/L_{\rm{Edd}}$ is the effective Eddington ratio. $A$ is a free parameter, and we need to find the optimal values of $A$ for given models. The relationship between $\theta_{\rm{0}}$ and $L_{\rm{bol}}/L_{\rm{Edd}}$ with $A=23$ is shown in Fig.~\ref{Fig2}, which suggests a positive relation between $\theta_{\rm{0}}$ and $L_{\rm{bol}}/L_{\rm{Edd}}$.

(2) The CF of dusty torus can be estimated by the fraction of type I AGNs. fig. 14 in \cite{Lu2010} presents the observational results of the relationship between the fraction of type I AGNs ($f_{\rm{1}}$) and ${\rm{Log}}(L_{\rm{bol}}/L_{\rm{Edd}})$. Thus we can obtain the relation between $\theta_{\rm{0}}$ and $L_{\rm{bol}}/L_{\rm{Edd}}$ by
\begin{equation}
f_{\rm{1}}=1-\cos \theta_{\rm{0}},
\end{equation}
where the Eddington luminosity $L_{\rm{Edd}}=\rm{1.3\times10^{38}}\left(\frac{\mathnormal{M}}{\mathrm{M}_{\bigodot}}\right)\ {\rm{erg\ s^{-1}}}$. There is a positive relation between $\theta_{\rm{0}}$ and $L_{\rm{bol}}/L_{\rm{Edd}}$ in this case, shown in Fig.~\ref{Fig2}.

(3) \cite{Dor2012} presented the numerical simulations on dusty tori. They took into account the pressure of infrared radiation on dust grains and the interaction of X-rays from a central black hole with hot and cold plasma, and determined the relation between X-ray luminosities and the critical angle of tori. Fig.~\ref{Fig2} shows the relation between the critical angle of tori and Eddington ratio from \cite{Dor2012}, which suggests an anti-correlation between $\theta_{\rm{0}}$ and $L_{\rm{bol}}/L_{\rm{Edd}}$. $\theta_{\rm{0}}$ is calculated by linear interpolating, while  $\theta_0=71^\circ$ for sources with ${\rm{Log}}(L_{\rm{bol}}/L_{\rm{Edd}})>-0.22$.

\subsection{The CF of the hot dust}
The hot dust CF is defined as the ratio of the thermal radiation ($L_{\rm{HD}}$) of hot dust to the bolometric luminosity ($L_{\rm{bol}}$) of accretion disc \citep{Mor2011}, i.e.,
\begin{equation}
{\rm{CF_{\rm{HD}}}}=\frac{L_{\rm{HD}}}{L_{\rm{bol}}},
\end{equation}
where $L_{\rm{bol}}=C'L_{\rm{uv}}$.
Generally, $C'=\rm{3\sim4.2}$ \citep{Mor2011}. In our work, we define $C'=\rm{3.3}$.

We assume that the dust is heated solely by exposure to the primary ultraviolet continuum and it is optically thin to the  reprocessed infrared continuum.

The calculation of $L_{\rm{HD}}$ is very similar to the process in \cite{Barvainis1987}. Therefore, we only briefly list the procedures. Please see the detailed process in the Section II of \cite{Barvainis1987}.

For the smooth case, the luminosity of hot dust is given by the integral over the volume $V$ of the torus defined by the critical angle discussed in Section 2.2,
\begin{equation}
L_{\rm{HD}}=\iiint n_{\rm{gr}}(r)L_{\rm{ir}}^{\rm{gr}} {\rm{d}} V\ \rm{erg\ s^{-1}},
\end{equation}
where $n_{\rm{gr}}$ is the number density of dusty grains, as described in Section 2.1, and $L_{\rm{ir}}^{\rm{gr}}=\int L_{\rm{\nu,ir}}^{\rm{gr}} {\rm{d}}\nu$ is the total luminosity of a dust grain, where $L_{\rm{\nu,ir}}^{\rm{gr}}$ is given by
\begin{equation}
L_{\rm{\nu,ir}}^{\rm{gr}}=4\uppi a^2\uppi Q_{\rm{\nu}}B_{\rm{\nu}}(T_{\rm{gr}})\ {\rm{erg\ s^{-1}\ {Hz}^{-1}}},
\end{equation}
where $Q_{\rm{\nu}}$ is the absorption efficiency of the grains in frequency $\rm{\nu}$, approximating $Q_{\rm{\nu}}$ with a power-law in the infrared is $Q_{\rm{\nu}}=q_{\rm{ir}}\nu^\gamma$, with $q_{\rm{ir}}=\rm{1.4\times10^{-24}}$ and $\gamma=\rm{1.6}$, shown in \cite{Barvainis1987}. $a$ is the radius of dust grain, here $a$ is equal to $0.05\ {\mu m}$. $T_{\rm{gr}}$ is the grain temperature, which is given by
\begin{equation}
T_{\rm{gr}}={\rm{1605}}(\frac{L_{\rm{uv,46}}f(\theta)}{r_{\rm{pc}}^{\rm{2}}})^{\rm{\frac{1}{5.6}}}e^{-\frac{\tau_{\rm{uv}}}{\rm{5.6}}}\ \rm{K},
\end{equation}
where $L_{\rm{uv,46}}$ is the ultraviolet luminosity in units of $\rm{10^{46}}$, and $r_{\rm{pc}}$ is the distance from the central source in parsecs. $\tau_{\rm{uv}}$ is the dust optical depth to the ultraviolet continuum, which is given by $\tau_{\rm{uv}}(r)=\uppi a^{\rm{2}}\int_{r_{\rm{1}}}^r n_{\rm{gr}}(r') {\rm{d}} r'$. The evaporation radius is
\begin{equation}
r_{\rm{1}}={\rm{1.2}}(L_{\rm{uv,46}}f(\theta))^{\rm{1/2}}T_{\rm{1500}}^{\rm{-2.8}}\ \rm{pc},
\end{equation}
where $T_{\rm{1500}}$ is the grain evaporation temperature in units of 1500 K. In \cite{Barvainis1987}, the radiation of accretion disc is assumed to be isotropic, i.e., $f(\theta)=1$. In this work, we also consider the anisotropic case, i.e., $f(\theta)=2\cos \theta$ ($\theta\subset\left[\theta_{\rm{0}},\frac{\pi}{\rm{2}}\right]$).

For the clumpy case, the dust is clumped into discrete clouds, which are optically thick to the ultraviolet continuum. Following the method in \cite{Barvainis1987}, the luminosity of the hot dust is given by
\begin{equation}
L_{\rm{HD}}=\iiint n_{\rm{cl}}(r)L_{\rm{ir}}^{\rm{cl}}e^{-\tau_{\rm{eff}}} {\rm{d}} V\ \rm{erg\ s^{-1}},
\end{equation}
where $n_{\rm{cl}}$ is the number density of dusty clouds, as described in Section 2.1. The luminosity of an individual cloud in the infrared  ($L_{\rm{ir}}^{\rm{cl}}$) is given by
\begin{equation}
L_{\rm{ir}}^{\rm{cl}}=\uppi l^{\rm{2}}n_{\rm{gr}}\int\limits_{\nu} \int_0^{x_{\rm{2}}} L_{\rm{\nu,ir}}^{\rm{gr}} {\rm{d}} x {\rm{d}} \nu\ \rm{erg\ s^{-1}},
\end{equation}
where $l$ is the radius of clumps, defining $x$ as the coordinate into the cloud along the direction from the central source, $x_{\rm{2}}$ is the depth at which $\tau_{\rm{uv}}=\rm{13}$.
 $\tau_{\rm{eff}}$ is an effective optical depth accounting for the occultation of different clumps,
\begin{equation}
\tau_{\rm{eff}}(r)=\uppi l^{\rm{2}}\int_{r_{\rm{1}}}^r n_{\rm{cl}}(r') {\rm{d}} r'=\pi N_{\rm{col,cl}}\int_{r_{\rm{1}}}^r \left(\frac{r'}{r_{\rm{0}}}\right)^{-\alpha} {\rm{d}} r'.
\end{equation}

The free parameters of all  models (Model $A$--$H$) are listed and explained in Table~\ref{Tab1}.

\section{Results}
Combining the assumptions in Section 2, there will be different models for $L_{\rm{HD}}$. Since equation (3) applies only to the anisotropic case and the relation of $\theta_{\rm{0}}\sim L_{\rm{bol}}/L_{\rm{Edd}}$, given by \cite{Dor2012}, is only applicable to the isotropic case, there are eight kinds of models in total. In order to find the optimal values of free parameters, we define ${\rm{rms}}=\sum_{i=1}^n ({\rm{CF}}_{{\rm{theory}},i}-{\rm{CF}}_{{\rm{obs}},i})^2$, where ${\rm{CF}}_{{\rm{theory}},i}$ is the value of theoretical CF for the $i$th source (i.e. CF calculated by model), and ${\rm{CF}}_{{\rm{obs}},i}$ is the observed CF of the $i$th source. The optimal values for free parameters are the values when $\rm{rms}$ is minimum.
The best-fitting values of parameters and the minimum values of $\rm{rms}$ are given in Table~\ref{Tab1}.  For each model, we show a  contour map of $\rm{rms}$ and  a contour map of the theoretical values of CF (i.e. ${\rm{CF}}_{\rm{theory}}$) with respect to $L_{\rm{bol}}$ and $M_{\rm{BH}}$  using the best-fitting parameters. A comparison of ${\rm{CF}}_{\rm{theory}}$ with ${\rm{CF}}_{\rm{obs}}$ is also shown to present the scatter of  data points  around the line of ${\rm{CF}}_{\rm{theory}}={\rm{CF}}_{\rm{obs}}$. Table~\ref{Tab1} also indicates the number of these figures for each model.

\begin{table*}
\centering
\begin{minipage}{180mm}
\caption{The models in our work.  $LU$, $LIU$ and $DOR$ mean the relation of $\theta_0\sim L_{\rm{bol}}/L_{\rm{Edd}}$ is given by Lu et al. (2010), Liu \& Zhang (2011), and Dorodnitsyn \&  Kallman (2012), respectively. $N_{\rm{col,cl}}$ is the column density of clouds with $N_{\rm{col,cl}}=n_{\rm{cl,0}}\times l^2$. $\alpha$ is the power-law index of the dust distribution. $A$ is the boosting factor of the effective cross-section due to the presence of dust (Liu \& Zhang 2011).}\label{Tab1}
\scalebox{0.9}{
\begin{tabular}{|c|c|c|c|r|r|r|c|c|}
\hline
 & & & & \multicolumn{3}{c|}{Free parameters} & &\\ \cline{5-7}
 &Dust Distribution&$\theta_0\sim L_{\rm{bol}}/L_{\rm{Edd}}$&$f(\theta)$&$N_{\rm{col,cl}}\ [\rm{pc^{-1}}]$&$\alpha$&$A$&rms&Results\\
\hline
Model $A$&CLUMPY&$LU$&$2\cos\theta$&$7.0\pm1.2$&$3.20\pm0.16$&$-$&120.23&Fig. 3\\
\hline
Model $B$&CLUMPY&$LU$&1&$2.04\pm0.19$&$2.55\pm0.08$&$-$&116.94&Fig. 4\\
\hline
Model $C$&CLUMPY&$LIU$&$2\cos\theta$&$1.14\pm0.23$&$2.325\pm0.022$&$0.26\pm0.26$&116.34&Fig. 5\\
\hline
Model $D$&CLUMPY&$DOR$&1&$33.4\pm0.3$&$2.91\pm0.29$&$-$&163.55&Fig. 6\\
\hline
Model $E$&SMOOTH&$LU$&$2\cos\theta$&$-$&$-$&$-$&155.88&Fig. 7\\
\hline
Model $F$&SMOOTH&$LU$&1&$-$&$-$&$-$&256.07&Fig. 8\\
\hline
Model $G$&SMOOTH&$LIU$&$2\cos\theta$&$-$&$-$&$4.69\pm0.07$&229.11&Fig. 9\\
\hline
Model $H$&SMOOTH&$DOR$&1&$-$&$-$&$-$&163.67&Fig. 10\\
\hline
\end{tabular}}
\end{minipage}
\end{table*}

\begin{figure}
\centering
\subfigure[]
{
\label{fig3:subfig:a}
\includegraphics[width=8cm]{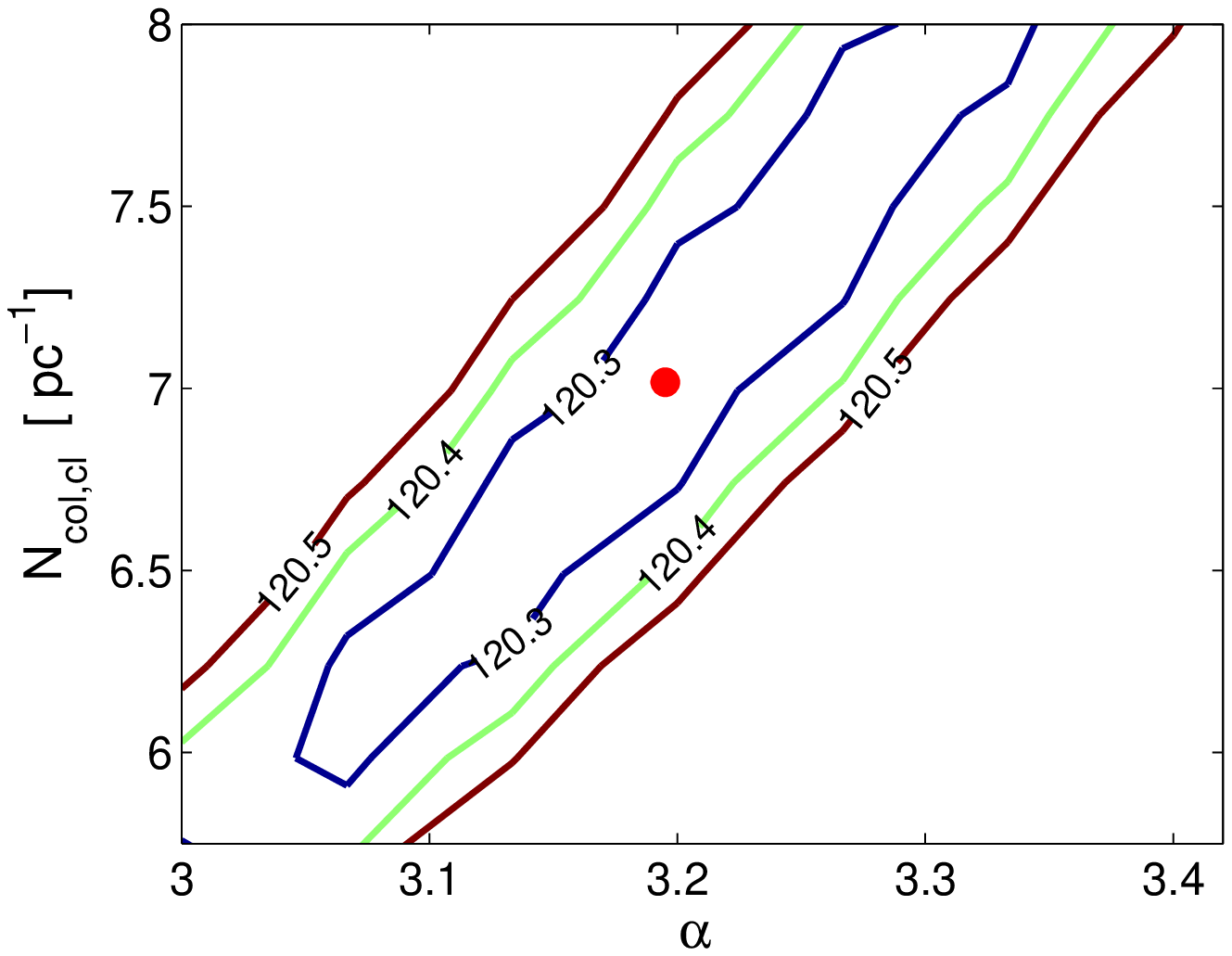}
}
\subfigure[]
{
 \label{fig3:subfig:b}
   \includegraphics[width=8cm]{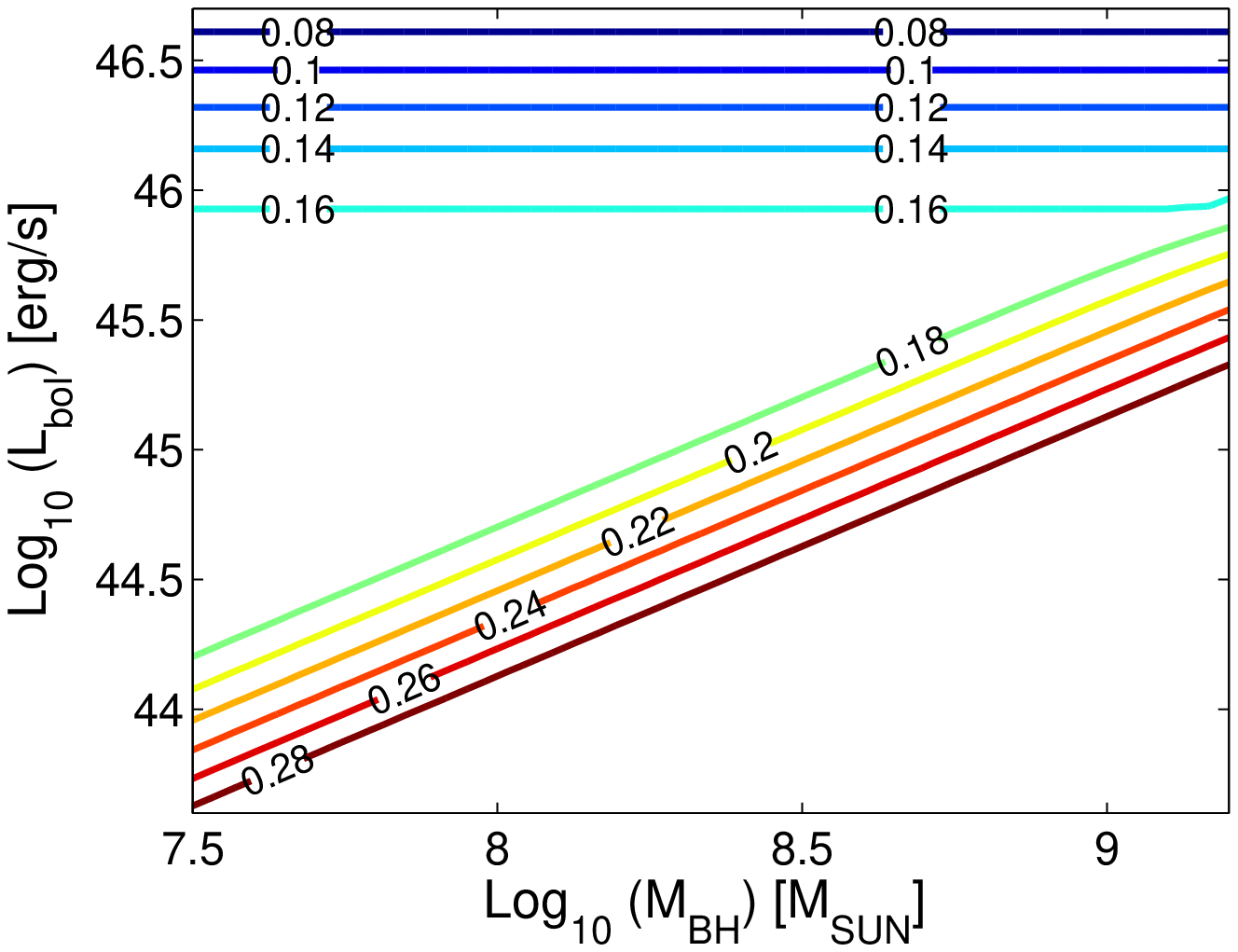}
}
\subfigure[]
{
 \label{fig3:subfig:c}
   \includegraphics[width=8cm]{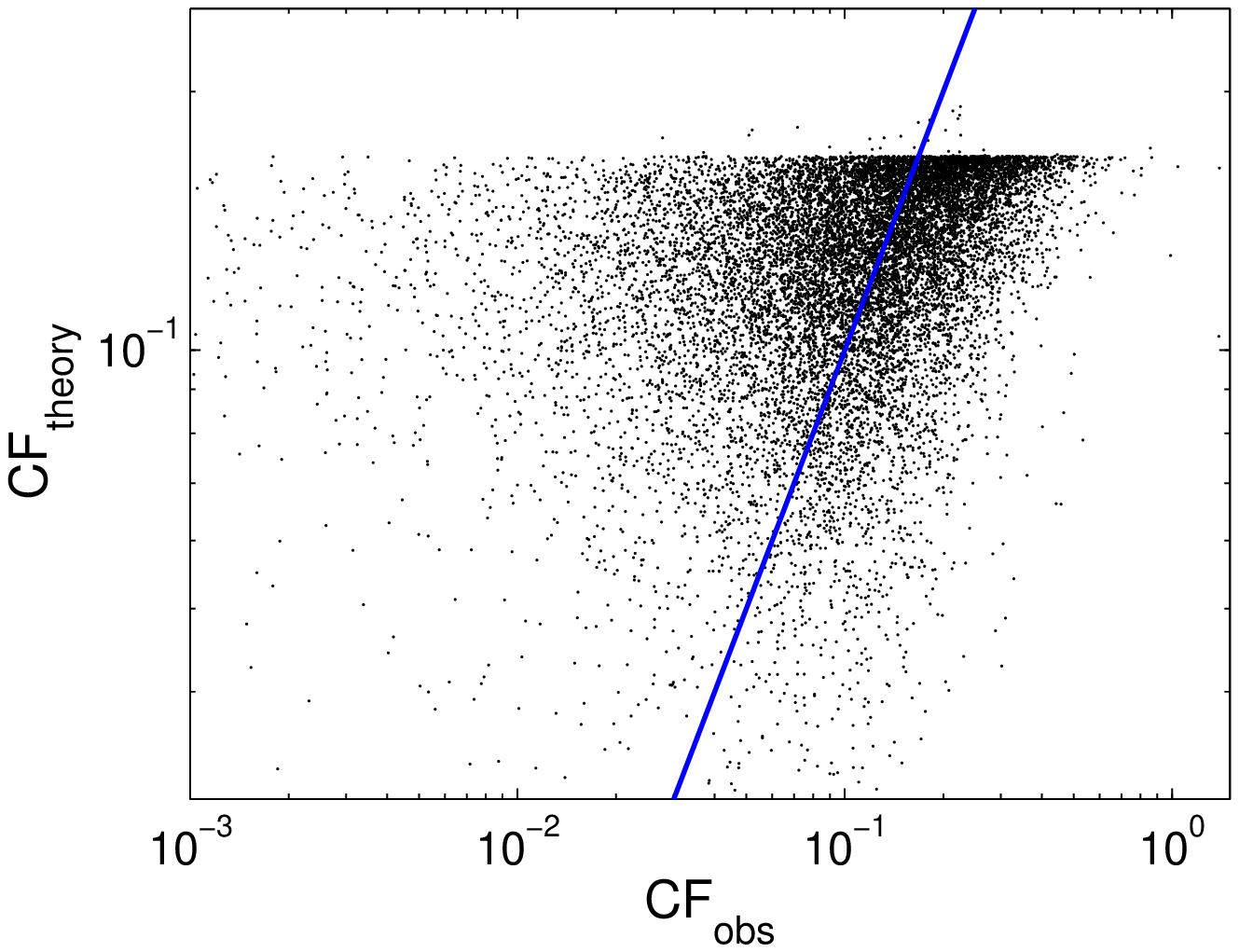}
}
\caption{The results of Model $A$. (a) The contour map of rms. Red dot is the best-fitting values of $N_{\rm{col,cl}}$ and $\alpha$. (b) The contour map of the theoretical values (i.e. ${\rm{CF}}_{\rm{theory}}$) with respect to $L_{\rm{bol}}$ and $M_{\rm{BH}}$ using the best-fitting parameters. (c) Comparison of the theoretical values (i.e. ${\rm{CF}}_{\rm{theory}}$) with the observed values (i.e. ${\rm{CF}}_{\rm{obs}}$): black dots are ${\rm{CF}}_{\rm{theory}}$ of all sources, blue solid line is the case of ${\rm{CF}}_{\rm{theory}}={\rm{CF}}_{\rm{obs}}$.}
\label{fig3:subfig}
\end{figure}

\begin{figure}
\centering
\subfigure[]
{
\label{fig4:subfig:a}
\includegraphics[width=8cm]{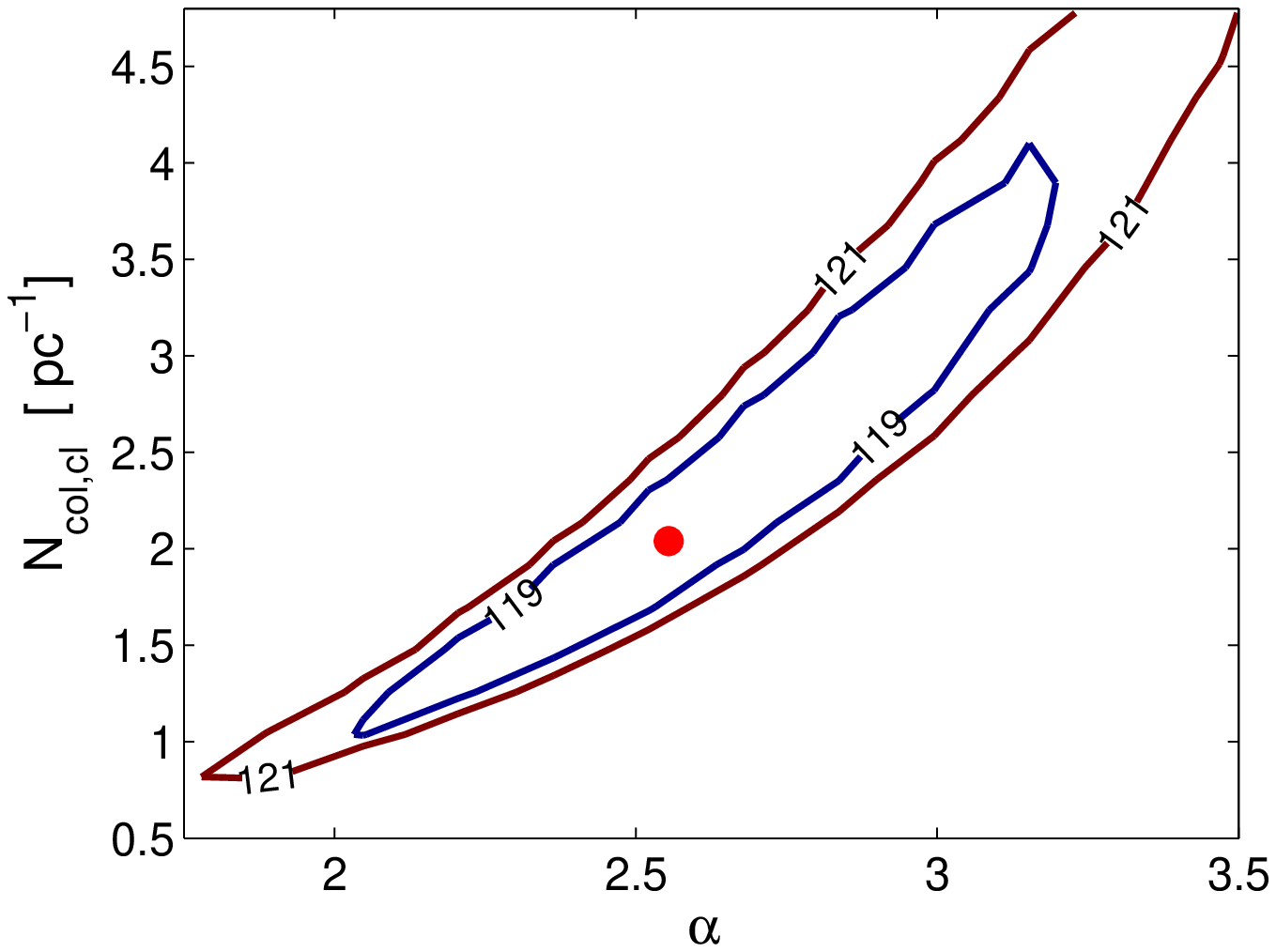}
}
\subfigure[]
{
 \label{fig4:subfig:b}
   \includegraphics[width=8cm]{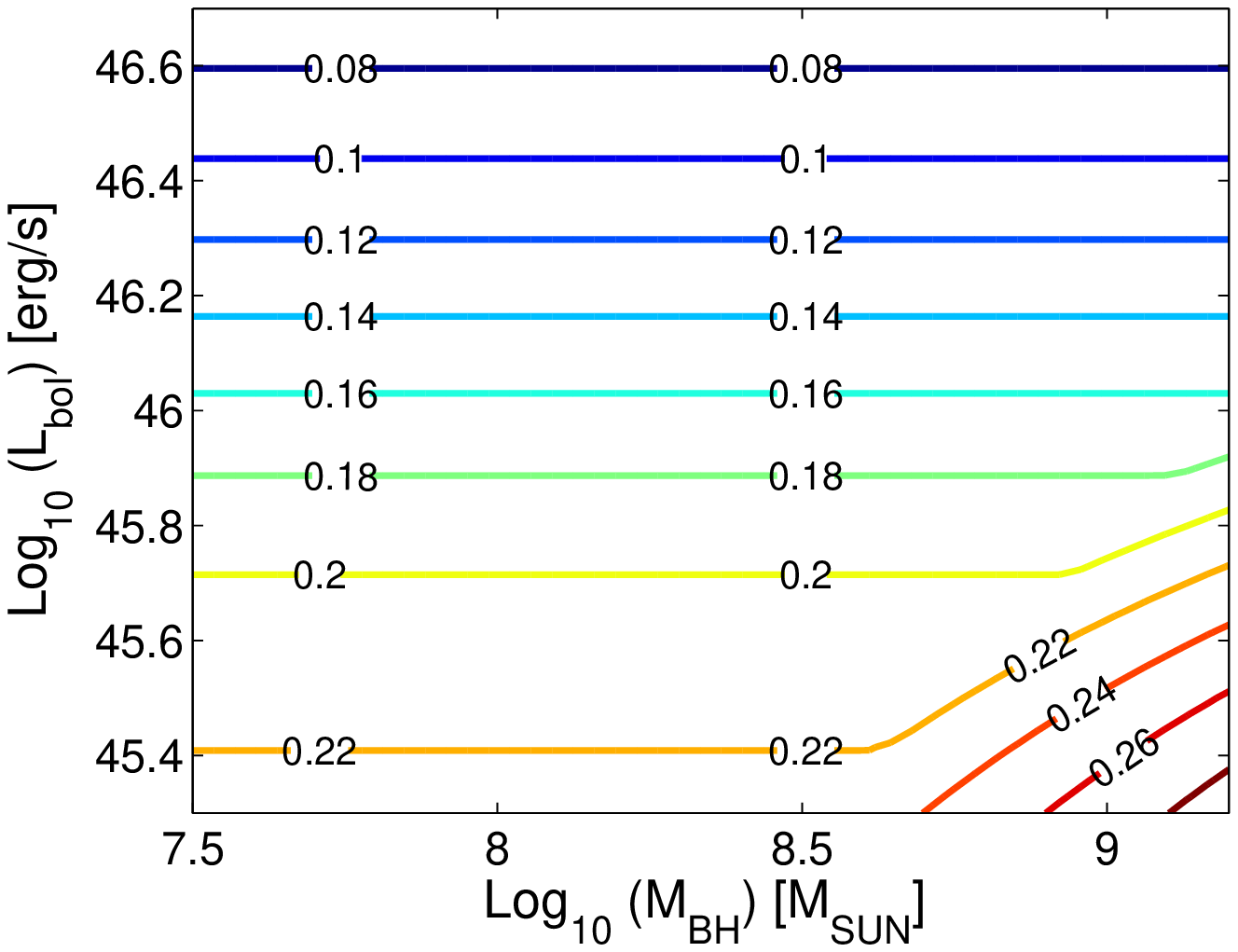}
}
\subfigure[]
{
 \label{fig4:subfig:c}
   \includegraphics[width=8cm]{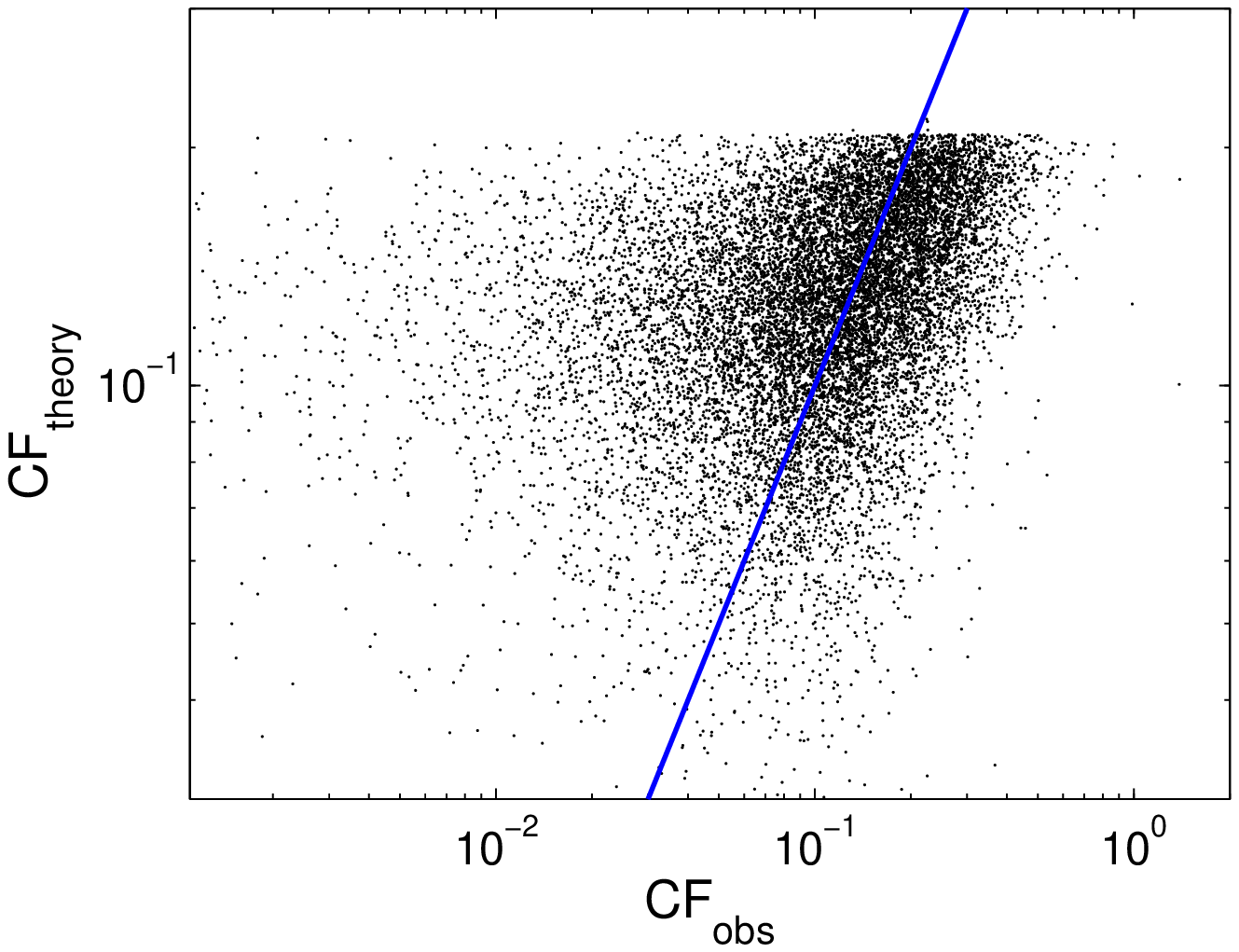}
}
\caption{The results of Model $B$. (a) The contour map of rms. Red dot is the best-fitting values of $N_{\rm{col,cl}}$ and $\alpha$. (b) The contour map of the theoretical values (i.e. ${\rm{CF}}_{\rm{theory}}$) with respect to $L_{\rm{bol}}$ and $M_{\rm{BH}}$ using the best-fitting parameters. (c) Comparison of the theoretical values (i.e. ${\rm{CF}}_{\rm{theory}}$) with the observed values (i.e. ${\rm{CF}}_{\rm{obs}}$): black dots are ${\rm{CF}}_{\rm{theory}}$ of all sources, blue solid line is the case of ${\rm{CF}}_{\rm{theory}}={\rm{CF}}_{\rm{obs}}$.}
\label{fig4:subfig}
\end{figure}

\begin{figure}
\centering
\subfigure[]
{
\label{fig5:subfig:a}
\includegraphics[width=7.5cm]{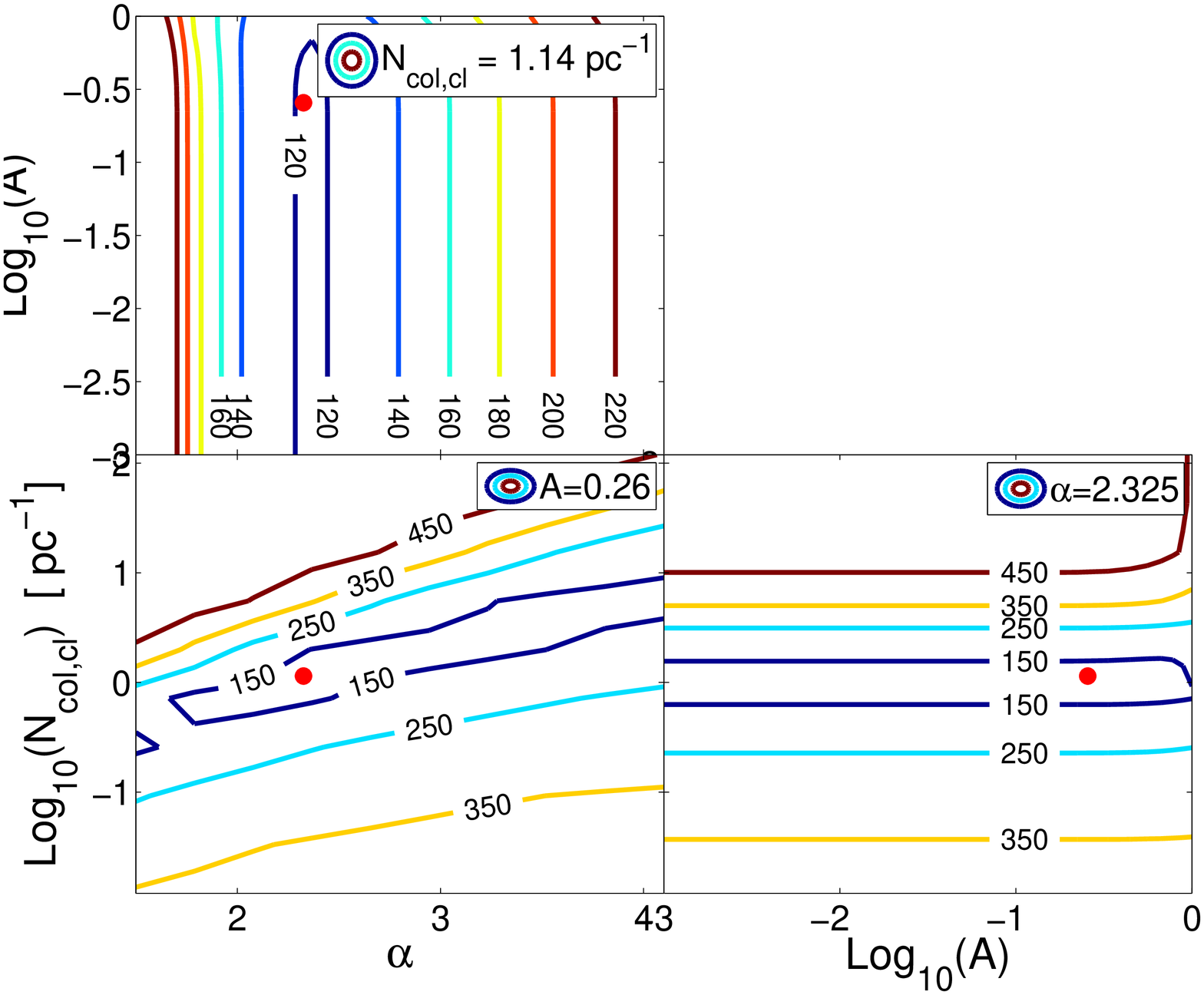}
}
\subfigure[]
{
 \label{fig5:subfig:b}
   \includegraphics[width=7.5cm]{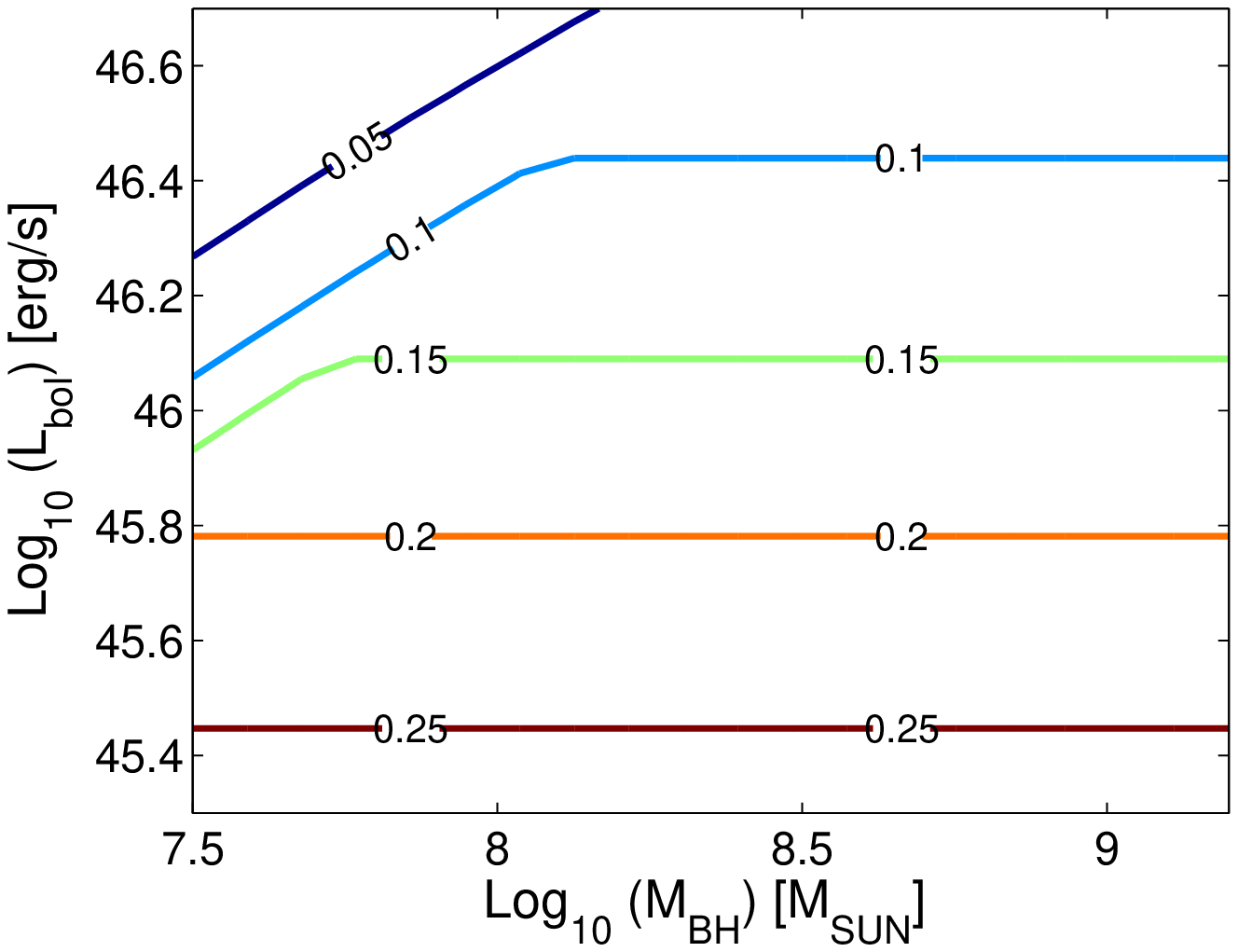}
}
\subfigure[]
{
 \label{fig5:subfig:c}
   \includegraphics[width=7.5cm]{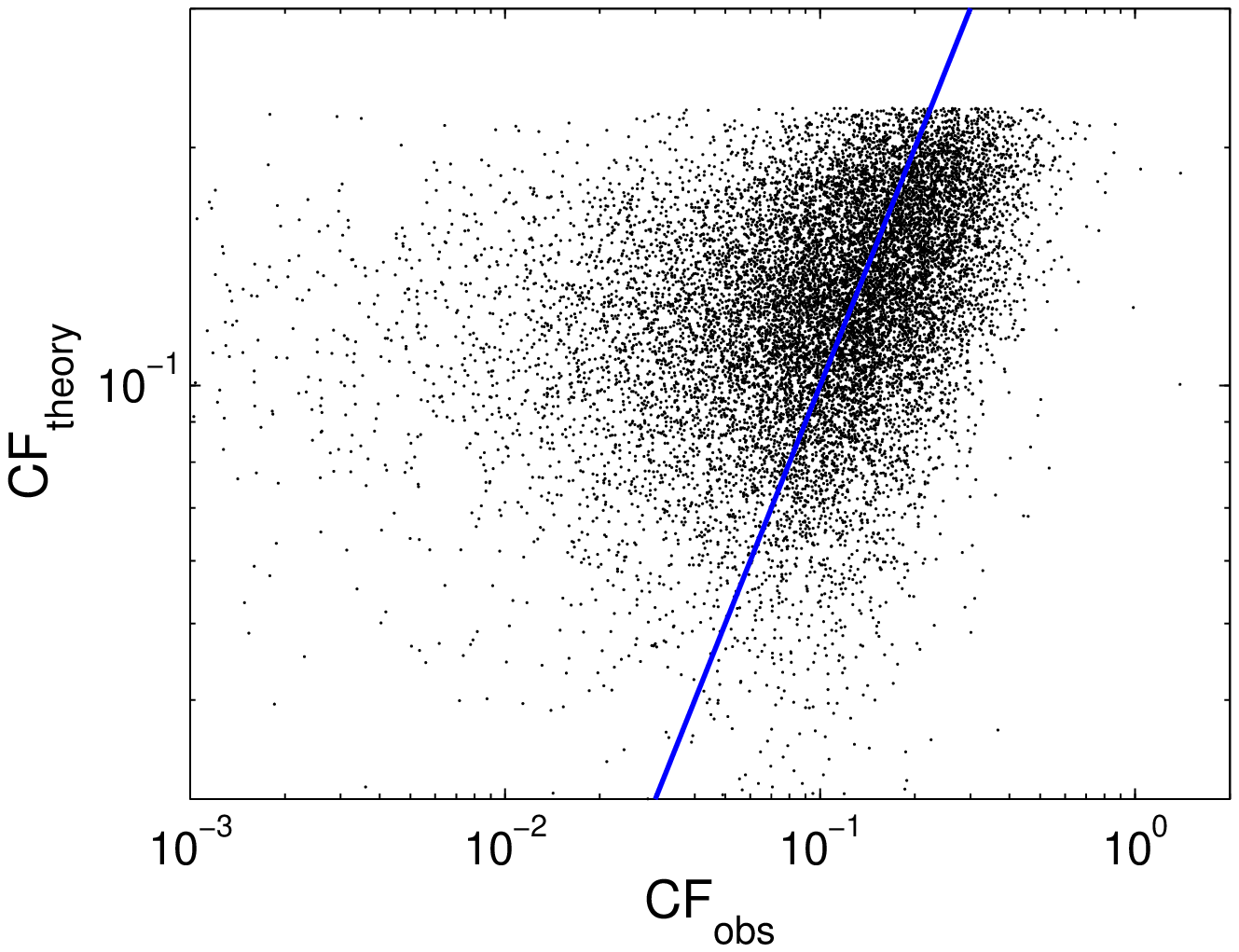}
}
\caption{The results of Model $C$. (a) The contour map of rms. The upper-left panel is the contour map of rms related to $A$ and $\alpha$, the lower-left panel is the contour map of rms related to $N_{\rm{col,cl}}$ and $\alpha$, and the lower-right panel is the contour map of rms related to $N_{\rm{col,cl}}$ and $A$. Red dots are the best-fitting values of $N_{\rm{col,cl}}$, $\alpha$ and $A$. The critical angle $\theta_0$ will be about $0^{\circ}$ with the case of $A<1$, which means the restriction of $A$ will be poor when $A<1$, just presented in the lower-right panel. (b) The contour map of the theoretical values (i.e. ${\rm{CF}}_{\rm{theory}}$) with respect to $L_{\rm{bol}}$ and $M_{\rm{BH}}$ using the best-fitting parameters. (c) Comparison of the theoretical values (i.e. ${\rm{CF}}_{\rm{theory}}$) with the observed values (i.e. ${\rm{CF}}_{\rm{obs}}$): black dots are ${\rm{CF}}_{\rm{theory}}$ of all sources, blue solid line is the case of ${\rm{CF}}_{\rm{theory}}={\rm{CF}}_{\rm{obs}}$.}
\label{fig5:subfig}
\end{figure}

\begin{figure}
\centering
\subfigure[]
{
\label{fig6:subfig:a}
\includegraphics[width=8cm]{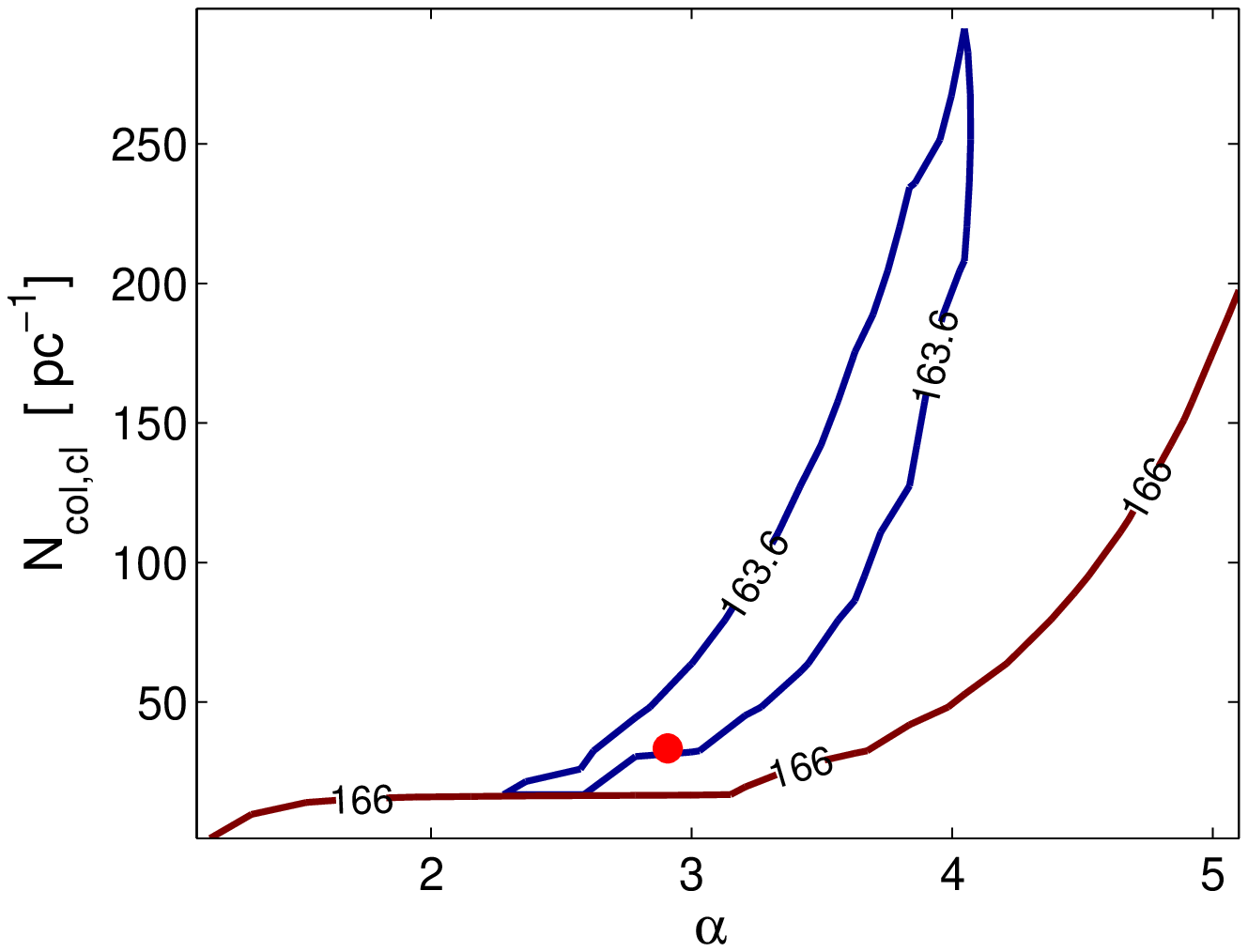}
}
\subfigure[]
{
 \label{fig6:subfig:b}
   \includegraphics[width=8cm]{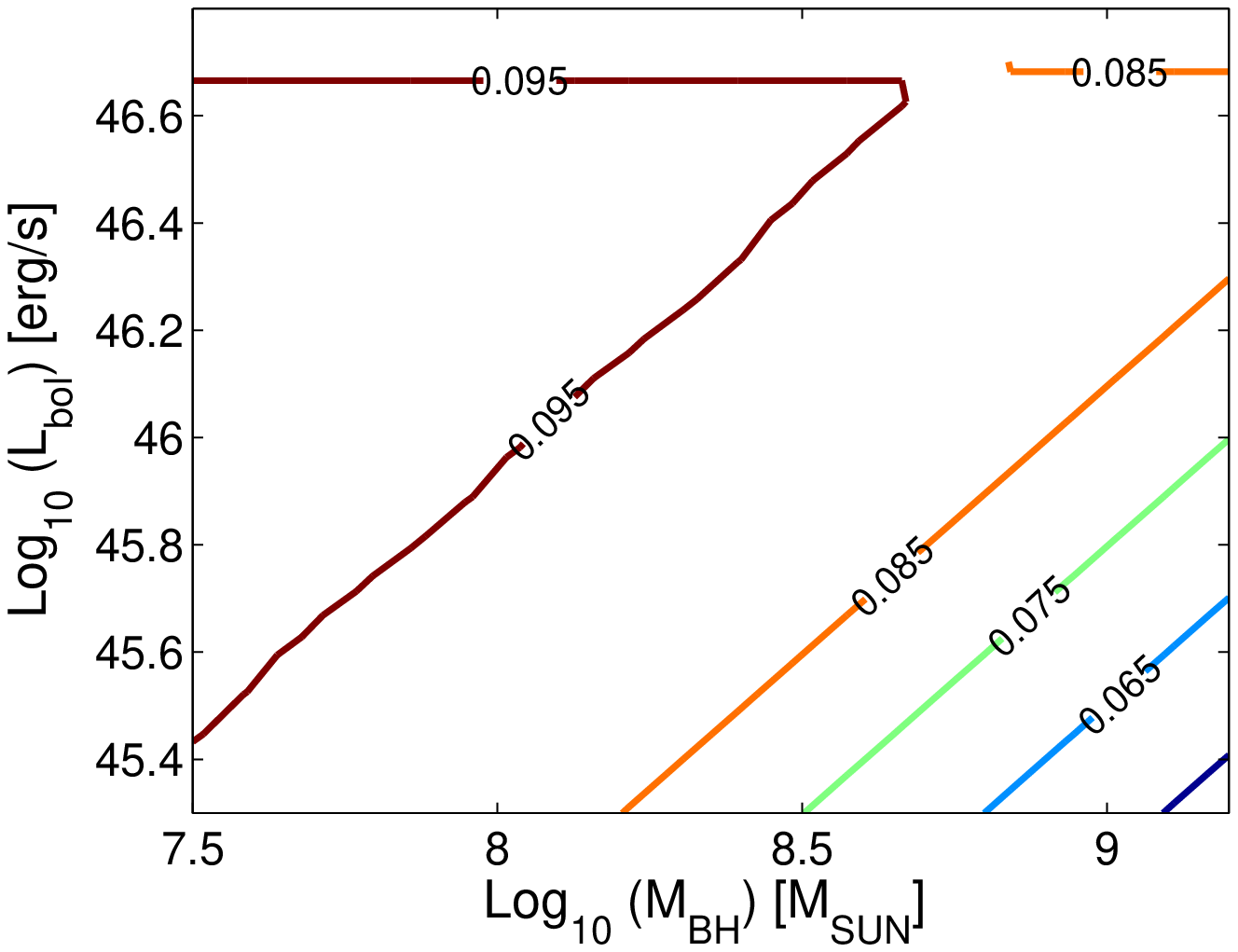}
}
\subfigure[]
{
 \label{fig6:subfig:c}
   \includegraphics[width=8cm]{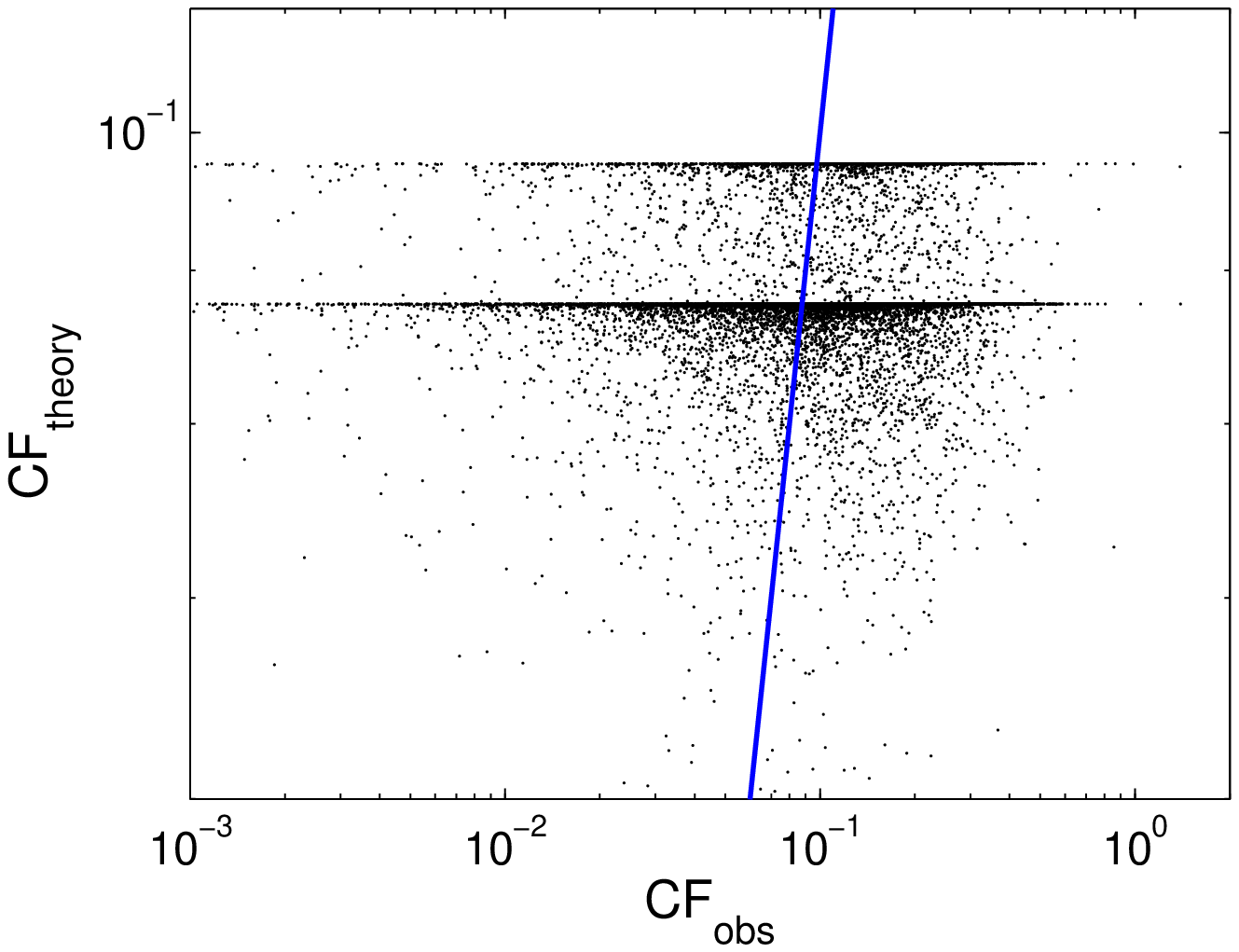}
}
\caption{The results of Model $D$. (a) The contour map of rms. Red dot is the best-fitting values of $N_{\rm{col,cl}}$ and $\alpha$. (b) The contour map of the theoretical values (i.e. ${\rm{CF}}_{\rm{theory}}$) with respect to $L_{\rm{bol}}$ and $M_{\rm{BH}}$ using the best-fitting parameters. (c) Comparison of the theoretical values (i.e. ${\rm{CF}}_{\rm{theory}}$) with the observed values (i.e. ${\rm{CF}}_{\rm{obs}}$): black dots are ${\rm{CF}}_{\rm{theory}}$ of all sources, blue solid line is the case of ${\rm{CF}}_{\rm{theory}}={\rm{CF}}_{\rm{obs}}$.}
\label{fig6:subfig}
\end{figure}

The clumpy structure enables the CF to be different from the geometry envelop of tori. Thus, the clumpy models (Models $A$--$C$), shown in Figs~\ref{fig3:subfig:c}, \ref{fig4:subfig:c} and \ref{fig5:subfig:c}, can generally produce the  required ${\rm{CF}}_{\rm{obs}}$ though with some scatters, except for Model $D$ (i.e. Fig.~\ref{fig6:subfig:c}). The anticorrelations between $\theta_0$ and $L_{\rm{bol}}/L_{\rm{Edd}}$ in Model $D$ results in a weak positive correlation between ${\rm{CF}}_{\rm{theory}}$ and $L_{\rm{bol}}$, which cannot present the observed anti-correlations between ${\rm{CF}}_{\rm{HD}}$ and $L_{\rm{bol}}$.

\begin{figure}
\centering
\subfigure[]
{
\label{fig7:subfig:a}
\includegraphics[width=8cm]{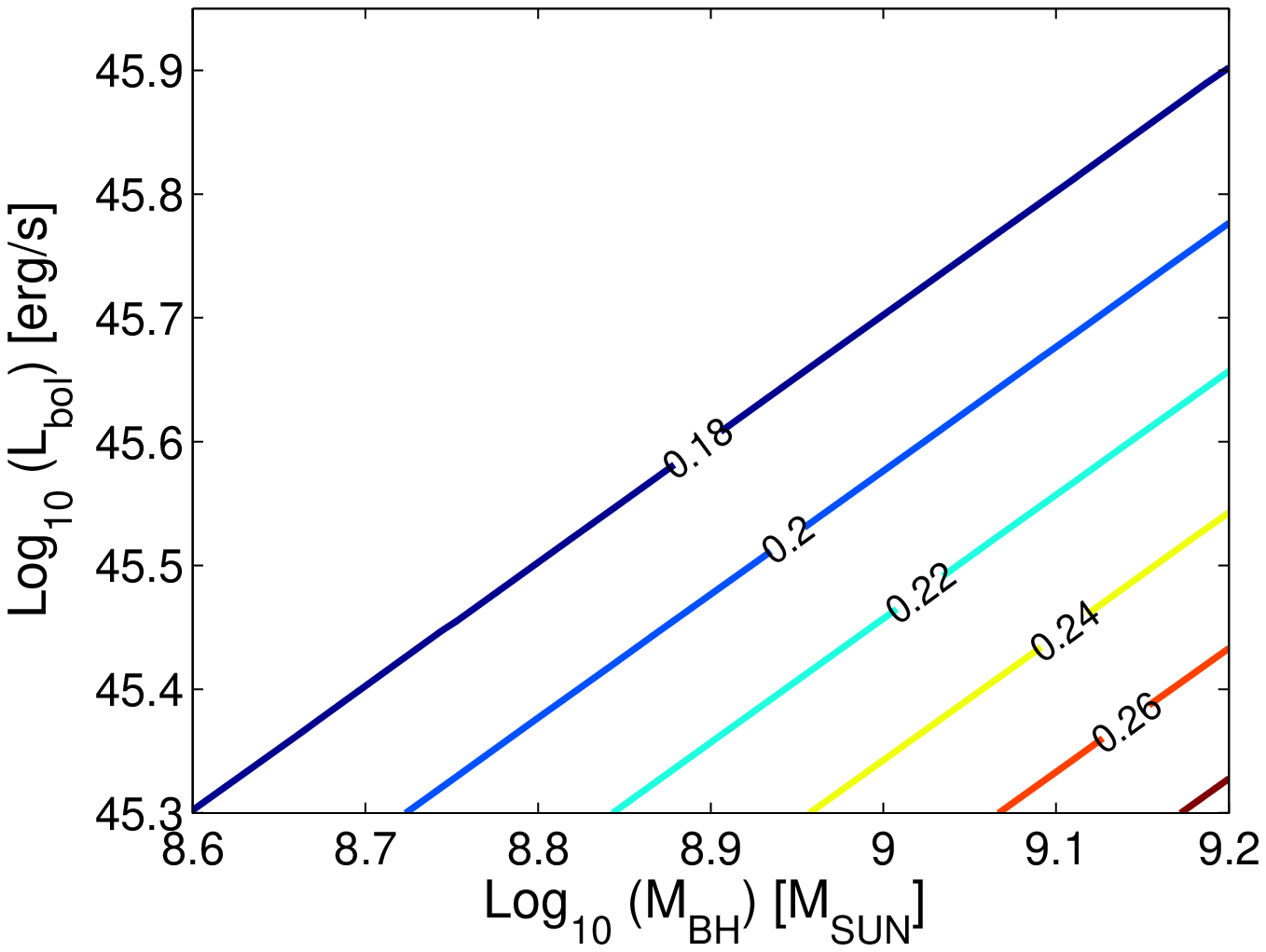}
}
\subfigure[]
{
 \label{fig7:subfig:b}
   \includegraphics[width=8cm]{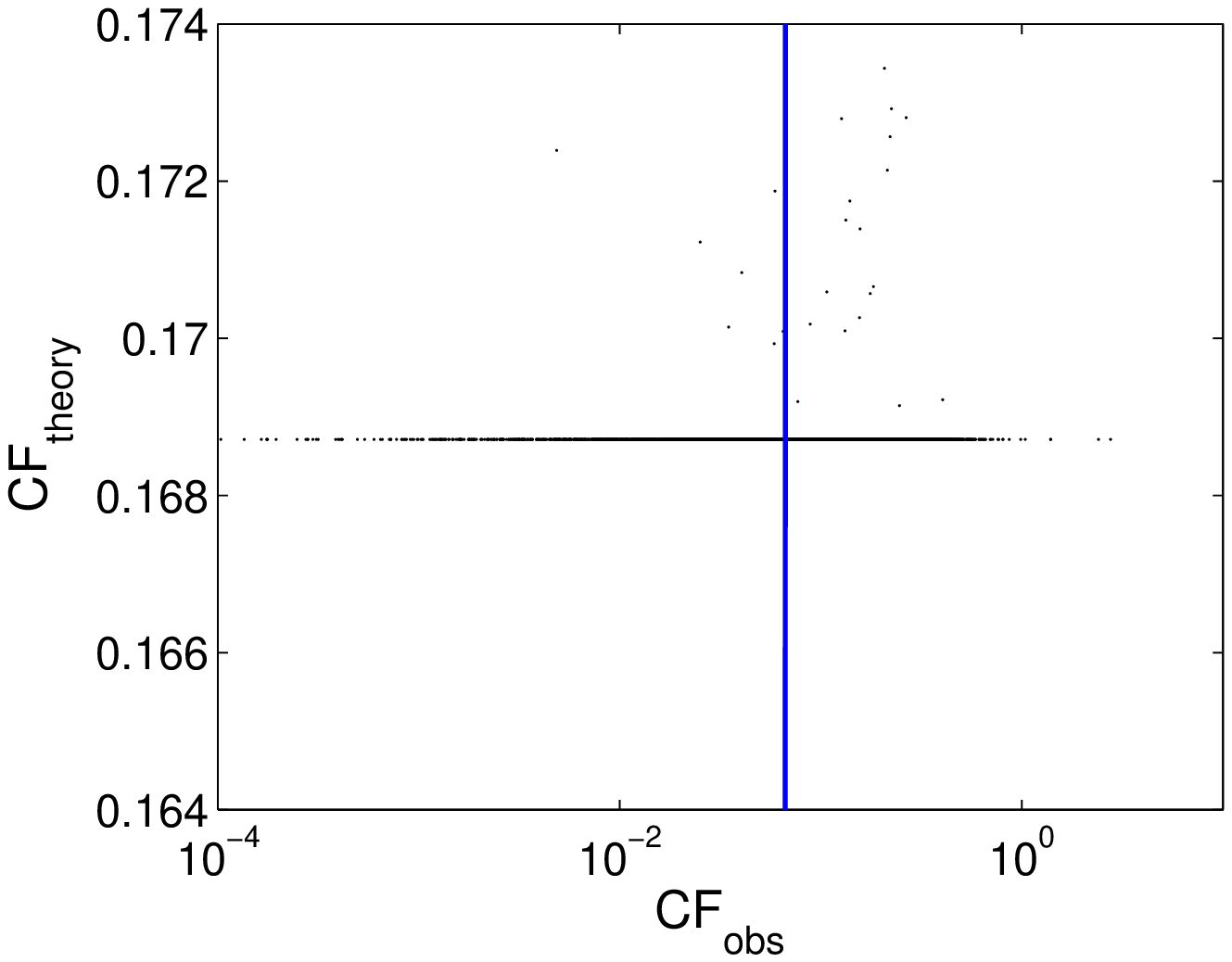}
}
\caption{The results of Model $E$. (a) The contour map of the theoretical values (i.e. ${\rm{CF}}_{\rm{theory}}$) with respect to $L_{\rm{bol}}$ and $M_{\rm{BH}}$ using the best-fitting parameters. (b) Comparison of the theoretical values (i.e. ${\rm{CF}}_{\rm{theory}}$) with the observed values (i.e. ${\rm{CF}}_{\rm{obs}}$): black dots are ${\rm{CF}}_{\rm{theory}}$ of all sources, blue solid line is the case of ${\rm{CF}}_{\rm{theory}}={\rm{CF}}_{\rm{obs}}$.}
\label{fig7:subfig}
\end{figure}

\begin{figure}
\centering
\subfigure[]
{
\label{fig8:subfig:a}
\includegraphics[width=8cm]{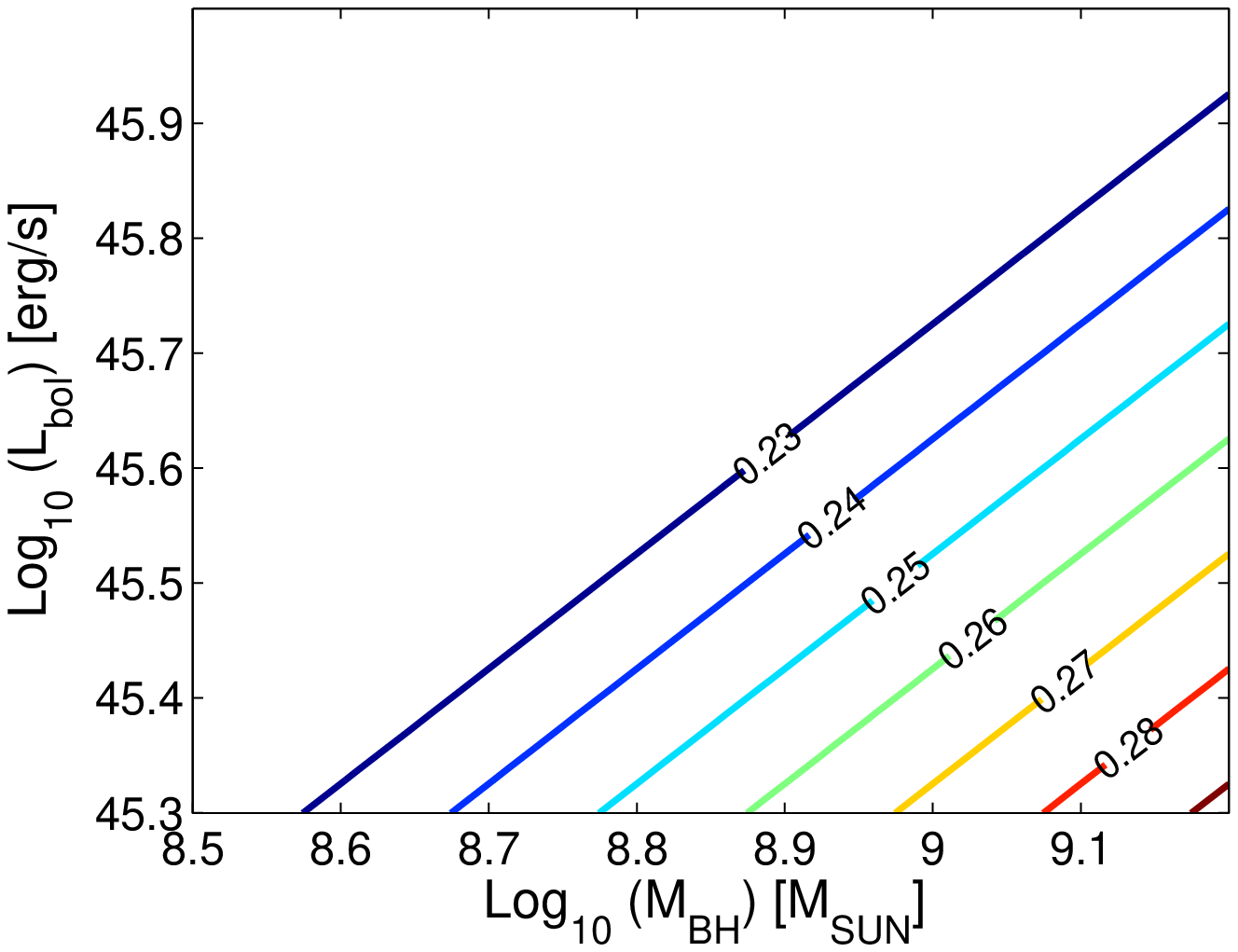}
}
\subfigure[]
{
 \label{fig8:subfig:b}
   \includegraphics[width=8cm]{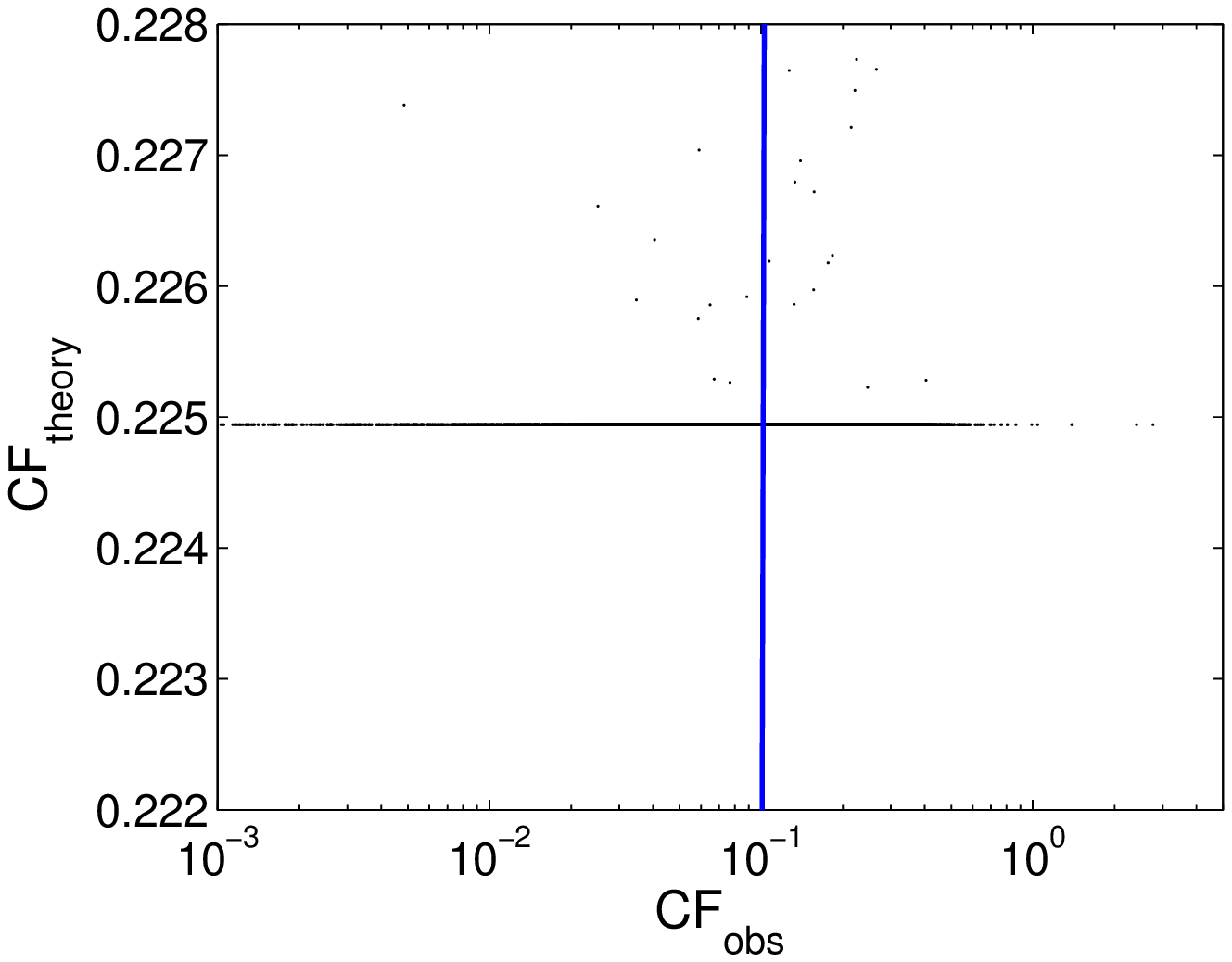}
}
\caption{The results of Model $F$. (a) The contour map of the theoretical values (i.e. ${\rm{CF}}_{\rm{theory}}$) with respect to $L_{\rm{bol}}$ and $M_{\rm{BH}}$ using the best-fitting parameters. (b) Comparison of the theoretical values (i.e. ${\rm{CF}}_{\rm{theory}}$) with the observed values (i.e. ${\rm{CF}}_{\rm{obs}}$): black dots are ${\rm{CF}}_{\rm{theory}}$ of all sources, blue solid line is the case of ${\rm{CF}}_{\rm{theory}}={\rm{CF}}_{\rm{obs}}$.}
\label{fig8:subfig}
\end{figure}

\begin{figure}
\centering
\subfigure[]
{
\label{fig9:subfig:a}
\includegraphics[width=8cm]{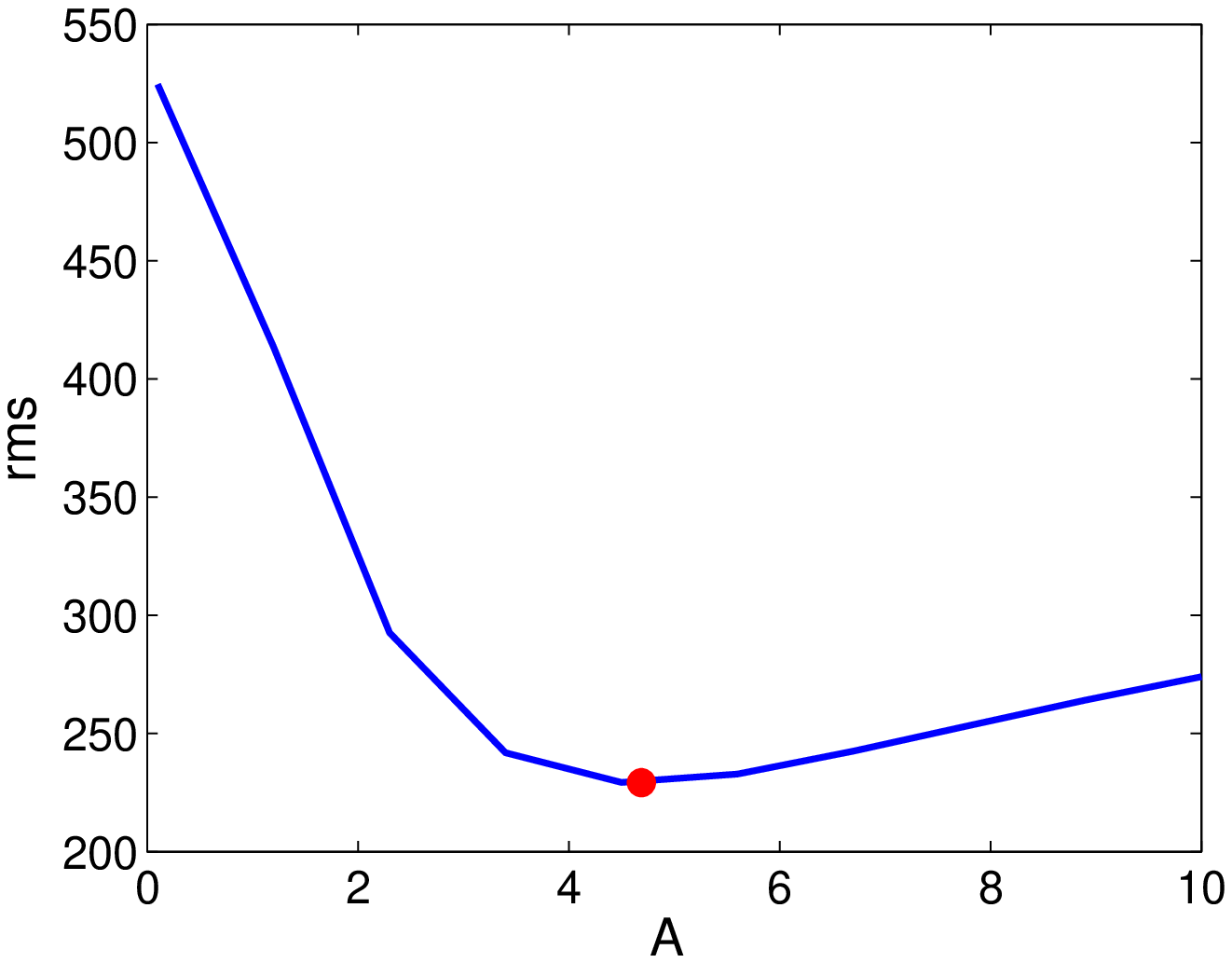}
}
\subfigure[]
{
 \label{fig9:subfig:b}
   \includegraphics[width=8cm]{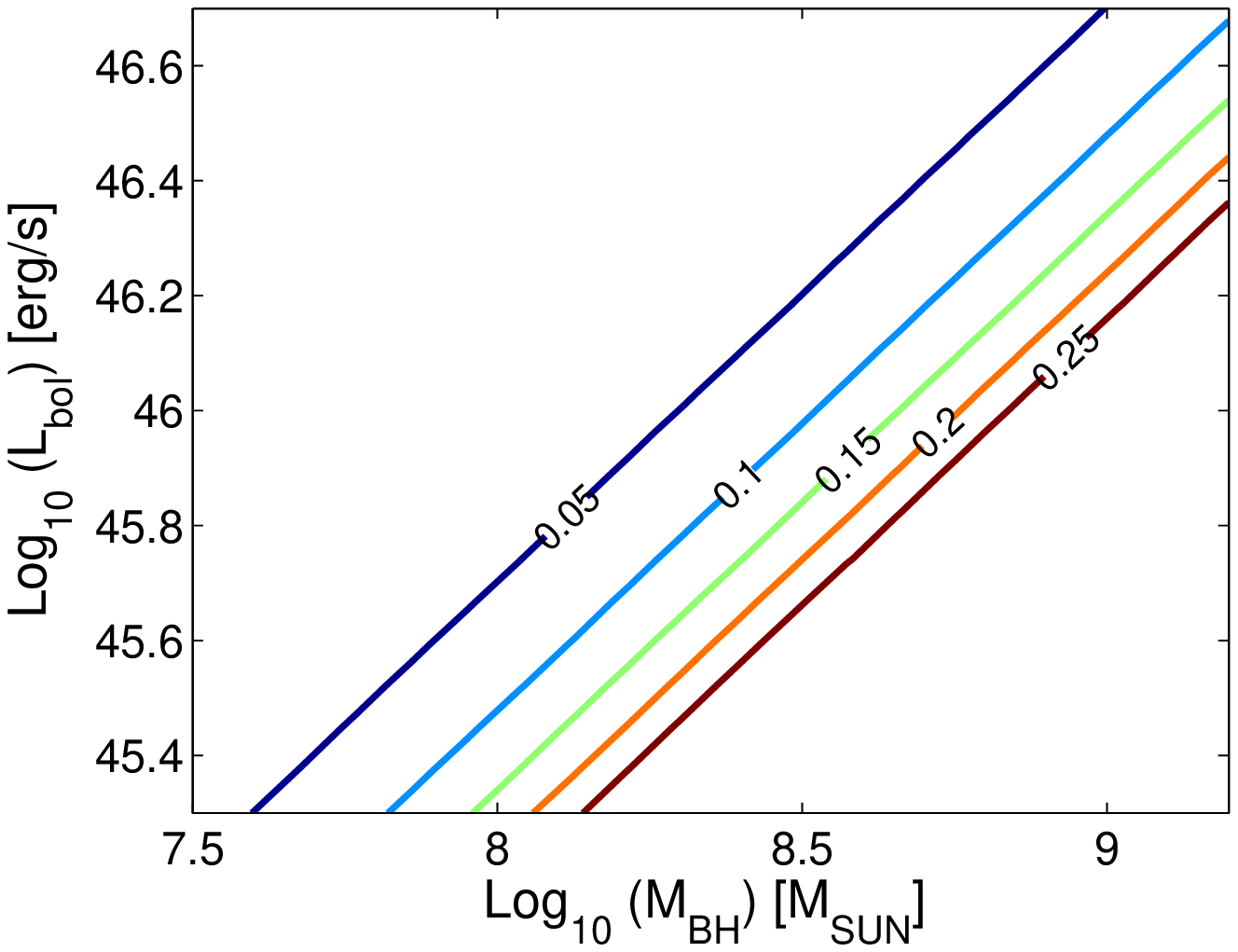}
}
\subfigure[]
{
 \label{fig9:subfig:c}
   \includegraphics[width=8cm]{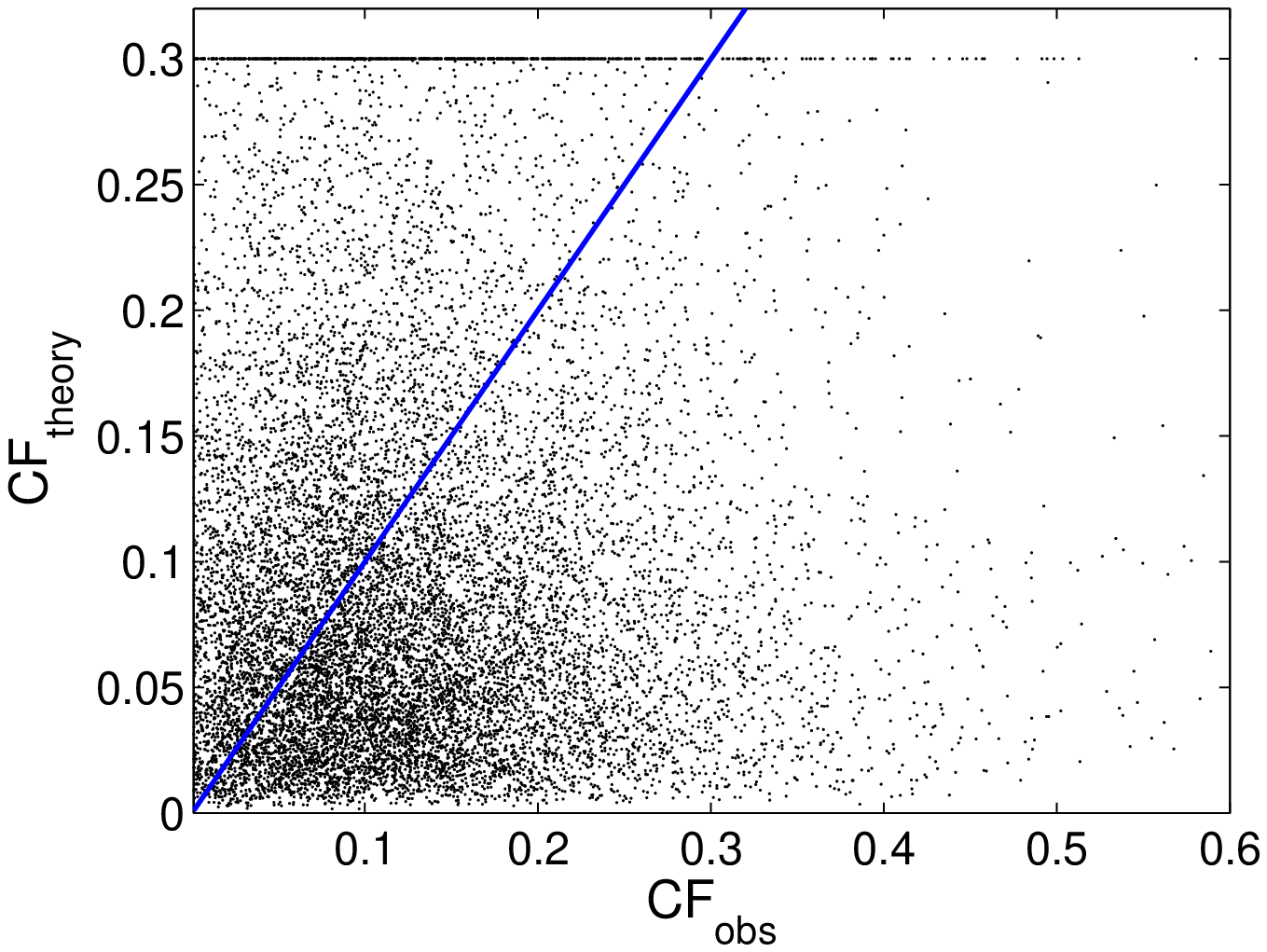}
}
\caption{The results of Model $G$. (a) The correlation between rms and $A$. Red dot is the best-fitting values of $A$. (b) The contour map of the theoretical values (i.e. ${\rm{CF}}_{\rm{theory}}$) with respect to $L_{\rm{bol}}$ and $M_{\rm{BH}}$ using the best-fitting parameters. (c) Comparison of the theoretical values (i.e. ${\rm{CF}}_{\rm{theory}}$) with the observed values (i.e. ${\rm{CF}}_{\rm{obs}}$): black dots are ${\rm{CF}}_{\rm{theory}}$ of all sources, blue solid line is the case of ${\rm{CF}}_{\rm{theory}}={\rm{CF}}_{\rm{obs}}$.}
\label{fig9:subfig}
\end{figure}

\begin{figure}
\centering
\subfigure[]
{
\label{fig10:subfig:a}
\includegraphics[width=8cm]{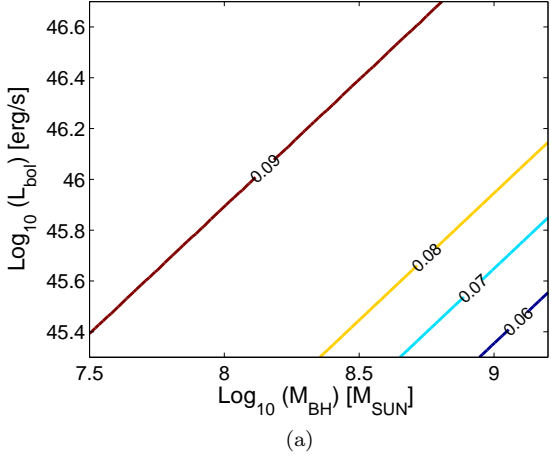}
}
\subfigure[]
{
 \label{fig10:subfig:b}
   \includegraphics[width=8cm]{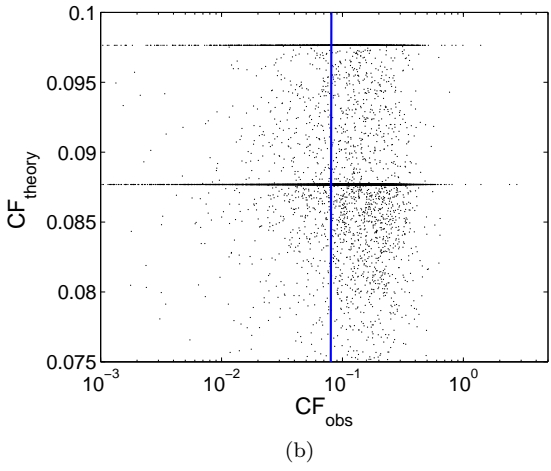}
}
\caption{The results of Model $H$. (a) The contour map of the theoretical values (i.e. ${\rm{CF}}_{\rm{theory}}$) with respect to $L_{\rm{bol}}$ and $M_{\rm{BH}}$ using the best-fitting parameters. (b) Comparison of the theoretical values (i.e. ${\rm{CF}}_{\rm{theory}}$) with the observed values (i.e. ${\rm{CF}}_{\rm{obs}}$): black dots are ${\rm{CF}}_{\rm{theory}}$ of all sources, blue solid line is the case of ${\rm{CF}}_{\rm{theory}}={\rm{CF}}_{\rm{obs}}$.}
\label{fig10:subfig}
\end{figure}

 In the case of smooth models (Models $E$--$H$), we fixed $n_{\rm{gr,0}}=10^{13.45}\ \rm{m^{-3}}$ and $\alpha=0.01$, since the optical depth of the torus is so large that the results are not sensitive to these parameters. As shown in Figs~\ref{fig7:subfig:b}, \ref{fig8:subfig:b}, \ref{fig9:subfig:c} and \ref{fig10:subfig:b}, the smooth models cannot reproduce the required ${\rm{CF}}_{\rm{obs}}$.

For Model $E$ and Model $F$, according to the relation between $\theta_{\rm{0}}$  and $L_{\rm{bol}}/L_{\rm{Edd}}$, the type I fraction is almost a constant with ${\rm{Log}}(L_{\rm{bol}}/L_{\rm{Edd}})>-1.5$, which is the most cases in our sample. Therefore,  the ${\rm{CF}}_{\rm{theory}}$ in Figs~\ref{fig7:subfig:b} and \ref{fig8:subfig:b} are almost the same values with different ${\rm{CF}}_{\rm{obs}}$.

For Model $G$, the free parameter $A$ makes the data points of ${\rm{CF}}_{\rm{theory}}$ scatter around the line of ${\rm{CF}}_{\rm{theory}}={\rm{CF}}_{\rm{obs}}$, shown in Fig.~\ref{fig9:subfig:c}.

For the relation between $\theta_{\rm{0}}$  and $L_{\rm{bol}}/L_{\rm{Edd}}$ of Model $H$,  the critical angles of tori are almost the constants if $-1.0<{\rm{Log}}(L_{\rm{bol}}/L_{\rm{Edd}})<-0.3$ or ${\rm{Log}}(L_{\rm{bol}}/L_{\rm{Edd}})>-0.22$. Since the two intervals include most of   the sources in our sample, the ${\rm{CF}}_{\rm{theory}}$ in Fig.~\ref{fig10:subfig:b} are concentrated at two values.

In order to quantify the above results, we use the Akaike information criterion ($AIC$) and the Bayesian information criterion ($BIC$) to select the best model from the eight models.

$AIC$ is a measure of the relative quality of a statistical model for a given set of data. For any statistical models, the $AIC$ value is given by
\begin{equation}
AIC=n{\rm{ln}}(\frac{\rm{rms}}{n})+2k+C,
\end{equation}
where $k$ is the number of free parameters in the model, $n$ denotes the sample size (in our sample, $n=13725$), ${\rm{rms}}=\sum_{i=1}^n ({\rm{CF}}_{{\rm{theory}},i}-{\rm{CF}}_{{\rm{obs}},i})^2$ is the residual sum of squares, which has been defined in section 3. $C$ can be ignored in model-comparisons. The best model is the model with the minimum value of $AIC$.

$BIC$ value is given by
\begin{equation}
BIC=n{\rm{ln}}(\hat{\sigma_e^2})+k{\rm{ln}}(n),
\end{equation}
where $\hat{\sigma_e^2}=\frac{1}{n}\sum_{i=1}^n (x_i-\overline{x})^2$ is the error variance, $x_i={\rm{CF}}_{{\rm{theory}},i}-{\rm{CF}}_{{\rm{obs}},i}$ and $\overline{x}=\frac{1}{n}\sum_{i=1}^n ({\rm{CF}}_{{\rm{theory}},i}-{\rm{CF}}_{{\rm{obs}},i})$. Similar to $AIC$, the best model is the model with the minimum $BIC$ value.

We define $\Delta AIC=AIC_{\rm{Model}\ X}-AIC_{\rm{Model}\ C}$, $\Delta BIC=BIC_{\rm{Model}\ X}-BIC_{\rm{Model}\ C}$, and $P_{\rm{AIC}}=e^{(AIC_{\rm{Model}\ C}-AIC_{\rm{Model}\ X})/2}$ \citep{Burnham2002}, where Model $X$ is one of the eight models in our work. The quantity $P_{\rm{AIC}}$ is the relative likelihood of Model $X$, which can be interpreted as the relative probability that the Model $X$ minimizes the information loss. The values of $\Delta AIC$, $\Delta BIC$ and $P_{\rm{AIC}}$ for all models are shown in Table~\ref{Tab2}.

Table~\ref{Tab2} clearly shows that the $AIC$ and $BIC$ value of Model $C$ are the minimum of all models. According to $AIC$ criteria, other models are nearly $P_{\rm{AIC}}=0$ times as probable as Model $C$ to minimize the information loss, and therefore the models, expect for Model $C$, will not be considered. According to $BIC$ criteria, Model $C$ will be better than Model $X$ very strongly when $\Delta BIC>10$ \citep{Kass1995}. Thus, Model $C$ is remarkably superior to other models according to $AIC$ and $BIC$ criterion.  In some models, the difference in rms is small, e.g. ${\rm{rms}}(\rm{Model}\ B)=116.94$ and ${\rm{rms}}(\rm{Model}\ C)=116.34$ (Table~\ref{Tab1}), while $\Delta AIC_{\rm{Model}\ B}=68$ and $\Delta BIC_{\rm{Model}\ B}=59$. The definitions of $AIC$ and $BIC$, i.e., equation (13) and (14), include the effect of sample size $n$ and the number of free parameters $k$. More specifically, the large sample size ($n=13725)$ has significantly amplified the difference in rms. Due to the lack of errors in the sample from \cite{Mor2011}, the $\chi^2$ statistics cannot be applied. However, if the errors of data are considered to be the same, the $AIC$ and $BIC$ criteria are equivalent to $\chi^2$ testing.

\section{CONCLUSIONS AND DISCUSSIONS}

 In this paper, we examined models to explain the observed correlations between the CF of host dust and the properties of AGNs found by  \cite{Mor2011}. Although the models adopted here are still not physical models, some heuristic assumptions explored here should be considered in further study.

Combining the possible dust distribution, the angular dependence of accretion disc radiation, and the critical angle of tori, we examined eight kinds of models. As a result, clumpy models (i.e. Models $A$--$C$) can explain the observed properties of tori, and all of smooth models significantly deviate from the observed CFs. Model $C$ is the best model selected with $AIC$ and $BIC$ criterion. The contour map in Fig.~\ref{fig5:subfig:b} indicates an anticorrelation between the CF and the bolometric luminosity and a weak correlation between the CF and the black hole mass. However, there is still obvious scatter around the required CF. We will explore if this is due to uncertainties in measuring $L_{\rm{bol}}$ and $M_{\rm{BH}}$.

\begin{table*}
\centering
\begin{minipage}{180mm}
\caption{The values of $\Delta AIC$ and $\Delta BIC$ for the models.}\label{Tab2}
\scalebox{0.9}{
\begin{tabular}{|c|c|c|c|c|}
\hline
 &$k$&$\Delta AIC$&$\Delta BIC$&${\rm{log_{10}}}(P_{\rm{AIC}})$\\
\hline
Model $A$&2&449&395&--97.5\\
\hline
Model $B$&2&68&59&--14.8\\
\hline
Model $C$&3&0&0&0\\
\hline
Model $D$&2&4668&2205&--1013.6\\
\hline
Model $E$&0&4009&2240&--870.5\\
\hline
Model $F$&0&10822&2239&--2350.0\\
\hline
Model $G$&1&9297&8378&--2018.8\\
\hline
Model $H$&0&4679&2314&--1016.0\\
\hline
\end{tabular}}
\end{minipage}
\end{table*}

\begin{figure}
\centering
\includegraphics[width=8cm]{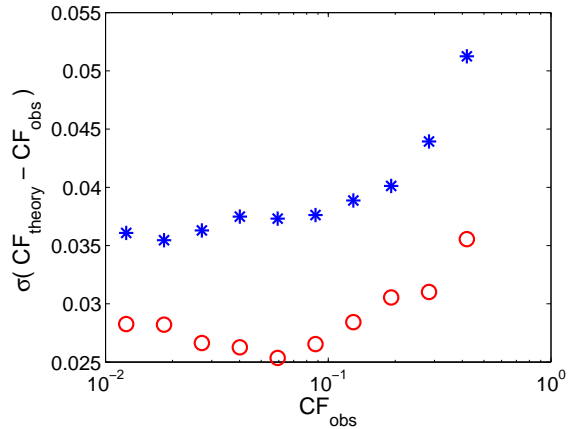}
\caption{The relations between $\sigma({\rm{CF}}_{\rm{theory}}-{\rm{CF}}_{\rm{obs}})$ and ${\rm{CF}}_{\rm{obs}}$. Blue stars: the observed scatter $\sigma({\rm{CF}}_{\rm{theory}}-{\rm{CF}}_{\rm{obs}})$ for Model $C$.  Red circles: the variance of ${\rm{CF}}_{\rm{theory}}$ for Model $C$ considering the errors of $L_{\rm{bol}}$ and $M_{\rm{BH}}$.}
\label{Fig32}
\end{figure}

For the observed decrease of ${\rm{CF}}_{\rm{HD}}$ with $L_{\rm{bol}}$ \citep{Mor2011}, $L_{\rm{bol}}$ is calculated from $L_{\rm{3000}}$ by applying a luminosity-dependant bolometric correction, which applied to the $SDSS/WISE$ sample range between about 3, for the most luminous sources, and 4.2, for the faintest \citep{Mor2011}. \cite{Trakhtenbrot2012} note that the real uncertainties on such estimates of $L_{\rm{bol}}$ are actually governed by the range of global SED variations between sources, as well as the assumed physical (or empirical) model for the UV SED. For example, the assumed exponent of the X-ray model and the $L_{\rm{X}}\sim L_{\rm{UV}}$ relation may only amount up to $\sim 0.2\ \rm{dex}$ in $L_{\rm{X}}$, and thus in the calculated $L_{\rm{bol}}$ \citep{Vignali2003,Bianchi2009}. On the other hand, for the uncertainties in measuring $M_{\rm{BH}}$, \cite{Feng2014} recently constructed three recipes for SE mass estimates and found the maximum scatter of  $M_{\rm{BH}}$ is $0.39\ \rm{dex}$. To determine the variance of ${\rm{CF}}_{\rm{theory}}$ in Model $C$ for the given uncertainties in  $L_{\rm{bol}}$ and $M_{\rm{BH}}$, we make a Gaussian random sampling of $M_{{\rm{BH}}}$ and $L_{{\rm{bol}}}$ of the sources in our sample with $\sigma_{\rm{M_{BH}}}=0.39\ \rm{dex}$ and $\sigma_{\rm{L_{bol}}}=0.2\ \rm{dex}$. This process is repeated for 1000 times and then the variance of ${\rm{CF}}_{\rm{theory}}$ is compared with $\sigma({\rm{CF}}_{\rm{theory}}-{\rm{CF}}_{\rm{obs}})=\sqrt{\hat{\sigma_e^2}}$ in Fig.~\ref{Fig32}. As the contribution of the uncertainties in measuring $L_{\rm{bol}}$ and $M_{\rm{BH}}$ to the scatter of data points is only about $75\%$, the uncertainties in  $L_{\rm{bol}}$ and $M_{\rm{BH}}$  cannot contribute to all the scatter.

Our results indicate $L_{\rm{bol}}$ and $M_{\rm{BH}}$ control the general structure of tori; however,
other factors, such as the hydrodynamics inside the torus, turbulent motions of these clumps \citep{Jaffe2004}, and the environment of the galaxies can also affect on the structure of torus, which should be included in the future models of tori. We can also examine the evolution of the dusty torus CF with redshift by the similar approach, which will be presented in future works.

\cite{Feltre2012} have performed a comparison between the smooth and clumpy models. They calculated several features of the IR model SEDs under smooth and clumpy configurations and found that the different behaviour of the silicate features at 9.7 and 18 $\upmu$m is mainly due to the different chemical compositions assumed by smooth and clumpy models. However, Feltre et al. (2012) also confirmed that the clumpy model will produce, on average, broader IR SEDs. The near-infrared index is also quite different between the two dust configurations as the result of different primary sources and the lack of a very hot component in the clumpy models. Their results indicate that the some features of the IR model SEDs may be not a clear diagnostic to distinguish between smooth and clumpy models. The models investigated in this paper are not the same as that adopted in \cite{Feltre2012}. In our models, the structure of the torus will change as the function of the black hole mass and luminosity, and it is shown that the responses of smooth and clumpy models are quite different. The clumpy torus models can generally explain the observed correlations of tori, while the smooth models fail to produce the required CFs. Therefore, the correlations between the CF of tori and the properties of AGNs provide a new tool to constrain the structure of tori.

\section*{Acknowledgements}
Here, we thank Rivay Mor and Benny Trakhtenbrot for providing the observed data sets of CF, $M_{\rm{BH}}$ and $L_{\rm{bol}}$ of all sources used in this work. This work is supported by 973 Program of China under grant
2014CB845802, by the National Natural Science Foundation of
China under grant nos. 11103019, 11133002, and 11103022, and 11373036, by the Qianren start-up grant 292012312D1117210, and by the Strategic Priority Research Program `The Emergence of Cosmological Structures' of the Chinese Academy of Sciences, Grant No. XDB09000000.

\bibliography{Ref}

\appendix
\section{}
\label{app}

To investigate the effect of the angular dependence of the density distribution of dust, a factor of $\sin\theta$ is included in the density distribution.
In the case of clumpy torus, the number density of dusty clumps is
\begin{equation}
n_{\rm{cl}}=n_{\rm{cl,0}}\left(\frac{r}{r_{\rm{0}}}\right)^{-\alpha}\sin\theta,
\end{equation}
and for the case of smooth torus, the number density of the dust is
\begin{equation}
n_{\rm{gr}}=n_{\rm{gr,0}}\left(\frac{r}{r_{\rm{0}}}\right)^{-\alpha}\sin\theta,
\end{equation}
where $\theta$ is the viewing angle of an observer (Fig.~\ref{Fig1}). As we will show, this specific and simple form is efficient to explain the effect of the angular dependence.

Following the same procedure in Section 3, we obtained the best-fitting values of free parameters and the minimum values of rms (Table~\ref{Tab 3}).  Compared with the results in Table~\ref{Tab1},  the column density of clouds slightly increases in the clumpy models (Models $A$--$D$). Since the clouds are concentrated towards the equatorial  plane, more clouds are required to produce the same CF. For the smooth models (Models $E$--$H$), since the optical depth of the torus is so large that the results are not sensitive to the angular dependence. More important, the minimum values of rms in all models are nearly identical to the cases without the factor of $\sin\theta$. Thus, the $AIC$ and $BIC$ value of Model $C$ are still the minimum of all models.

Although we have only adopted some particular forms of the angular dependence, it is clear that the form of angular dependence will only change the best-fitting values of free parameters, e.g., the numbers of clouds required to produce the observed CF. The intrinsic characteristics of the smooth and clumpy models will not change with the angular dependence, i.e. the general trend of the predicted correlations between CF and the properties of AGNs are not influenced by the specific form of the angular dependence. As a result, the relative relations of the $AIC$ and $BIC$ values of different models will not change. Thus, our main conclusion, the clumpy models explain the observed correlations of tori better than the smooth models, is not sensitive to the assumed angular dependence of the distribution of dust.

\begin{table*}
\centering
\begin{minipage}{180mm}
\caption{The results under the assumption that the density distribution of the dust depends on both the viewing angle $\theta$ and radius $r$.}\label{Tab 3}
\scalebox{0.9}{
\begin{tabular}{|c|c|c|c|r|r|r|c|c|c|c|c|}
\hline
 & & & & \multicolumn{3}{c|}{Free parameters} & & & & &\\ \cline{5-7}
 &Dust Distribution&$\theta_0\sim L_{\rm{bol}}/L_{\rm{Edd}}$&$f(\theta)$&$N_{\rm{col,cl}}\ [\rm{pc^{-1}}]$&$\alpha$&$A$&rms&$k$&$\Delta AIC$&$\Delta BIC$&${\rm{log_{10}}}(P_{\rm{AIC}})$\\
\hline
Model $A$&CLUMPY&$LU$&$2\cos\theta$&$10.15\pm0.06$&$3.34\pm0.04$&$-$&120.27&2&436&382&--94.7\\
\hline
Model $B$&CLUMPY&$LU$&1&$2.3\pm0.8$&$2.6\pm2.0$&$-$&116.95&2&52&43&--11.3\\
\hline
Model $C$&CLUMPY&$LIU$&$2\cos\theta$&$2.21\pm0.11$&$2.517\pm0.006$&$0.266\pm0.017$&116.48&3&0&0&0\\
\hline
Model $D$&CLUMPY&$DOR$&1&$34.5\pm0.4$&$2.916\pm0.019$&$-$&163.58&2&4658&2207&--1011.5\\
\hline
Model $E$&SMOOTH&$LU$&$2\cos\theta$&$-$&$-$&$-$&155.88&0&3992&2223&--866.9\\
\hline
Model $F$&SMOOTH&$LU$&1&$-$&$-$&$-$&256.07&0&10805&2222&--2346.3\\
\hline
Model $G$&SMOOTH&$LIU$&$2\cos\theta$&$-$&$-$&$4.69\pm0.07$&229.11&1&9280&8361&--2015.1\\
\hline
Model $H$&SMOOTH&$DOR$&1&$-$&$-$&$-$&163.67&0&4662&2297&--1012.3\\
\hline
\end{tabular}}
\end{minipage}
\end{table*}

\bibliographystyle{mn2e}
\end{document}